\documentclass[a4paper,twoside,10pt,english,numbers=noenddot]{scrbook}

\usepackage[english]{babel}
\makeatletter
\def\month@ngerman{\ifcase\month \or Januar\or Februar\or M\"arz\or April\or Mai\or Juni\or Juli\or August\or September\or Oktober\or November\or Dezember\fi}
\def\month@english{\ifcase\month \or January\or February\or March\or April\or May\or June\or July\or August\or September\or October\or November\or December\fi}
\makeatother
\usepackage{hyphenat} 

\usepackage[T1]{fontenc}
\usepackage{helvet} 
\usepackage{indentfirst}
\usepackage[keeplastbox]{flushend} 

\usepackage[usenames,dvipsnames]{xcolor} 
\usepackage{graphicx} 
\usepackage{wrapfig} 
\usepackage{tikz}
\usepackage{rotating}

\usepackage{ifthen}
\usepackage{forloop}
\usepackage{etoolbox}

\usepackage[fleqn]{amsmath} 
\usepackage{amssymb} 
\usepackage{amsfonts} 
\usepackage{amstext} 
\usepackage{wasysym}
\usepackage{mathrsfs} 
\usepackage{mathtools}
\usepackage{upgreek}
\usepackage{siunitx}
\usepackage{esint} 
\usepackage{cancel} 
\usepackage{empheq}
\allowdisplaybreaks

\usepackage{multicol} 
\usepackage{multirow} 
\usepackage{dcolumn}
\usepackage{tabularx} 
\newcolumntype{L}[1]{>{\raggedright\arraybackslash\hsize=#1\hsize}X}
\newcolumntype{R}[1]{>{\raggedleft\arraybackslash\hsize=#1\hsize}X}
\newcolumntype{C}[1]{>{\centering\arraybackslash\hsize=#1\hsize}X}
\usepackage{dblfloatfix} 

\usepackage{enumitem} 
\setlist{nosep} 

\usepackage[a4paper,portrait]{geometry}
\newlength{\TwoColumnWidth}
\newlength{\OneColumnWidth}
\usepackage[headsepline]{scrlayer-scrpage}

\usepackage[hang,bottom]{footmisc}
\renewcommand{\thefootnote}{$\ddagger$}
\deffootnote{\footnotemargin}{0pt}{%
	\textsuperscript{\thefootnotemark}
}
\usepackage[singlelinecheck=false,font=small,labelfont=bf,format=plain]{caption} 
\usepackage[singlelinecheck=true,font=small,labelfont=bf,format=plain]{subfig}
\setkomafont{sectioning}{\rmfamily\bfseries}
\setkomafont{descriptionlabel}{\rmfamily\bfseries}

\usepackage[subfigure]{tocloft} 
\addto\captionsenglish{

}
\makeatletter
  \renewcommand*{\@pnumwidth}{20pt} 
  \renewcommand*{\@tocrmarg}{30pt plus 5pt minus 0pt} 
  \renewcommand*{\@dotsep}{4} 
\makeatother

\newcounter{chapterappendixcounter}[chapter]

\newcounter{totalpagecounter}
\newcounter{totalfigurecounter}
\newcounter{totaltablecounter}
\newcounter{totalcitecounter}
\newcounter{totalpages}
\newcounter{totalfigures}
\newcounter{totaltables}
\newcounter{totalcites}
\makeatletter
\@input{\jobname.myaux}
\makeatother

\definecolor{white}{rgb}{1,1,1}
\definecolor{black}{rgb}{0,0,0}
\definecolor{red}{rgb}{1,0,0}
\definecolor{green}{rgb}{0,1,0}
\definecolor{blue}{rgb}{0,0,1}
\definecolor{cyan}{rgb}{0,1,1}
\definecolor{magenta}{rgb}{1,0,1}
\definecolor{yellow}{rgb}{1,1,0}
\definecolor{darkgreen}{rgb}{0,0.6,0}
\definecolor{darkyellow}{rgb}{0.8,0.8,0}
\definecolor{orange}{rgb}{1,0.5,0}
\definecolor{tuc}{RGB}{0,90,70}
\definecolor{tuclight}{RGB}{218,234,194}
\definecolor{tucorange}{RGB}{242,148,0}
\definecolor{tucbg}{RGB}{224,233,233}

\usepackage[
	colorlinks=true,                   
	urlcolor=blue,                     
	filecolor=blue,                    
	linkcolor=blue,                    
	citecolor=blue,                    
	breaklinks=true,                   
	backref=page,                      
	pagebackref=true,                  
	bookmarks=true,                    
	bookmarksopen=false,               
	bookmarksnumbered=true,            
	pdfauthor={Fabian Teichert},       
	pdfdisplaydoctitle=true,           
	pdfstartview=FitH,                 
	pdfpagelayout=TwoPageRight,        
]{hyperref}

\newif\ifsinglepaper\singlepaperfalse
\addto\captionsenglish{
	\renewcommand\bibname{References}
}

\renewcommand*{\backref}[1]{}
\renewcommand*{\backrefalt}[4]{%
	\ifsinglepaper\else%
		\ifcase #1
		\or (cited at p.~#2).
		\else (cited at pp.~#2).
		\fi%
	\fi
}
\usepackage{cite} 

\newcommand{\bstindent}{99}
\newcommand{\bstaddress}{}
\newcommand{\bstauthor}{}

\newcommand{\bstjournal}{}

\newcommand{\bstpublisher}{}

\newcommand{\bsttitle}{\itshape}
\newcommand{\bstvolume}{}
\newcommand{\bstyear}{}
\newcommand{\bbland}{and}

\newcommand{\bblnov}{November}

\newcommand\bibliographysection{\section}
\newcommand\bibliographysectionstyle{}
\newcommand\bibliographyitemsize{\normalsize}
\newcommand\bibliographyitemseparation{}
\makeatletter
\newcommand\bibcontentsline{\addcontentsline{toc}{section}{References}}
\renewenvironment{thebibliography}[1]{%
	\bibliographysection{\bibliographysectionstyle\bibname}
	\bibcontentsline%
	\renewcommand\rightmark{\bibname}
	\list{\@biblabel{\@arabic\c@enumiv}}{\settowidth\labelwidth{\@biblabel{#1}}%
		\leftmargin\labelwidth%
		\advance\leftmargin\labelsep%
		\@openbib@code%
		\usecounter{enumiv}%
		\let\p@enumiv\@empty%
		\renewcommand\theenumiv{\@arabic\c@enumiv}%
	}%
	\sloppy
	\clubpenalty4000
	\@clubpenalty \clubpenalty
	\widowpenalty4000%
	\sfcode`\.\@m%
}{%
	\def\@noitemerr{\@latex@warning{Empty `thebibliography' environment}}%
	\endlist%
}
\makeatother
\let\oldthebibliography\thebibliography
\renewcommand\thebibliography[1]{
	\bibliographyitemsize
	\oldthebibliography{#1}
	\bibliographyitemseparation
}

\let\oldtwocolumn\twocolumn
\let\oldonecolumn\onecolumn
\newif\iftwocolumn\twocolumntrue
\def\onecolumn{\twocolumnfalse}
\def\twocolumn{\twocolumntrue}
\newif\ifarticlestyle\articlestylefalse

\renewcommand{\title}[2]{%
	\setcounter{authors}{0}%
	\setcounter{addresses}{0}%
	\setcounter{keywords}{0}%
	\def\inserttitle{#1}%
	\def\articlelabel{#2}
}
\newcommand{\email}[1]{\def\insertemail{#1}}
\newcommand{\abstract}[1]{\def\insertabstract{#1}}
\def\insertjournal{}
\def\insertjournalshort{}
\def\insertdoi{}
\def\insertarxiv{}
\def\insertarxivshort{}
\newcommand{\journal}[4][nothing]{%
	\def\insertjournalshort{#2}%
	\if\relax\detokenize{#3}\relax%
		\def\insertjournal{}%
	\else%
		\def\tmpa{#1}%
		\def\tmpb{submitted}%
		\def\tmpc{accepted}%
		\ifx\tmpa\tmpb%
			\def\journalpre{Submitted to: }%
		\else%
			\ifx\tmpa\tmpc%
				\def\journalpre{Accepted in: }%
			\else%
				\def\journalpre{}%
			\fi%
		\fi%
		\if\relax\detokenize{#4}\relax\def\insertjournal{\journalpre#3}\else\def\insertjournal{\journalpre\href{#4}{#3}}\fi%
	\fi%
}
\newcommand{\doi}[1]{\if\relax\detokenize{#1}\relax\def\insertdoi{}\else\def\insertdoi{DOI: \href{http://dx.doi.org/#1}{#1}}\fi}
\newcommand{\arxiv}[2]{\if\relax\detokenize{#1}\relax\def\insertarxiv{}\def\insertarxivshort{}\else\def\insertarxiv{arXiv: \href{https://arxiv.org/abs/#1}{#1 [#2]}}\def\insertarxivshort{arXiv: #1}\fi}
\newcounter{authors}
\newcounter{addresses}
\newcounter{keywords}
\newcommand{\addauthor}[2]{\csdef{author\arabic{authors}}{#1}\csdef{authoraddress\arabic{authors}}{#2}\stepcounter{authors}}
\newcommand{\addaddress}[1]{\csdef{address\arabic{addresses}}{#1}\stepcounter{addresses}}
\newcommand{\addkeyword}[1]{\csdef{keyword\arabic{keywords}}{#1}\stepcounter{keywords}}

\newcounter{otherchapter}
\newcounter{normalchapter}
\newcounter{othercounter}
\newcounter{i}
\newcounter{j}
\makeatletter
\let\normalchapter\chapter

\renewcommand\chapter{%
	\@ifstar{%
		\normalchapter*%
	}{%
		\stepcounter{normalchapter}%
		\normalchapter%
	}%
}
\newcommand\otherchapter{%
	\protected@write\@auxout{}{\string\@writefile{lof}{\string\addvspace{10\string\p@}}}%
	\protected@write\@auxout{}{\string\@writefile{lot}{\string\addvspace{10\string\p@}}}%
	\scr@startsection{chapter}{1}{\z@}{0ex \@plus -0.2ex}{3.5ex \@plus 0.2ex}{\Large\bfseries}%
}
\newcommand\reftype{}
\newcommand\reflabel{}
\newcommand\refnumber{}
\newcommand\refshortnumber{}
\newcommand\reftext{}
\newcommand{\articletitlesub}{%
	\renewcommand\reftype{chapter}%
	\renewcommand\reflabel{section*.\arabic{othercounter}}%
	\renewcommand\refnumber{\Alph{otherchapter}}%
	\renewcommand\refshortnumber{}%
	\renewcommand\reftext{\inserttitle\ifx\insertjournalshort\empty\ (\insertarxivshort)\else\ (\insertjournalshort)\fi}%
	\pdfbookmark[0]{\reftext}{\reflabel}%
	\otherchapter*{\inserttitle}%
	\protected@write\@auxout{}{\string\@writefile{toc}{\string\contentsline {\reftype}{\string\numberline {\refnumber}\reftext}{\thepage}{\reflabel}}}%
	\Alabel{\articlelabel}%
	\noindent\textbf{%
		\large\csuse{author0}$^{\csuse{authoraddress0}}$%
		\forloop{i}{1}{\value{i} < \value{authors}}{%
			, \csuse{author\arabic{i}}$^{\csuse{authoraddress\arabic{i}}}$%
		}
	}\\[1em]
	\normalsize
	\setcounter{j}{0}
	\forloop{i}{0}{\value{i} < \value{addresses}}{%
		\stepcounter{j}
		$^{\arabic{j}}$\,\csuse{address\arabic{i}}
		\ifthenelse{\value{j}<\value{addresses}}{\\}{}
	}
	\ifx\insertemail\empty\\[1em]\else\\[0.5em]E-mail address: \insertemail\\[1em]\fi
	\textbf{Abstract:} \insertabstract
	\ifthenelse{\value{keywords}=0}{}{
		\\[1em]
		Keywords: \csuse{keyword0}%
			\forloop{i}{1}{\value{i} < \value{keywords}}{%
				; \csuse{keyword\arabic{i}}%
			}
	}
}
\newcommand{\articletitle}{%
	\setcounter{articlepage}{0}%
	\stepcounter{otherchapter}%
	\stepcounter{chapter}%
	\setcounter{section}{0}%
	\setcounter{subsection}{0}%
	\setcounter{subsubsection}{0}%
	\stepcounter{othercounter}%
	\iftwocolumn\oldtwocolumn[\articletitlesub\vspace{1.5em}]\else\oldonecolumn\articletitlesub\fi%
}%
\let\oldchapter\chapter
\let\oldsection\section
\let\oldsubsection\subsection
\let\oldsubsubsection\subsubsection
\newcommand\articlesectiondata[1]{%
	\renewcommand\reftype{section}%
	\renewcommand\reflabel{section.\arabic{chapter}.\arabic{section}}%
	\renewcommand\refnumber{\Alph{otherchapter}.\arabic{section}}%
	\renewcommand\refshortnumber{\arabic{section}}%
	\renewcommand\reftext{#1}%
}
\newcommand\articlesubsectiondata[1]{%
	\renewcommand\reftype{subsection}%
	\renewcommand\reflabel{subsection.\arabic{chapter}.\arabic{section}.\arabic{subsection}}%
	\renewcommand\refnumber{\Alph{otherchapter}.\arabic{section}.\arabic{subsection}}%
	\renewcommand\refshortnumber{\arabic{section}.\arabic{subsection}}%
	\renewcommand\reftext{#1}%
}
\newcommand\articlesectionnostar[1]{%
	\articlesectiondata{#1}%
	\pdfbookmark[1]{\reftext}{\reflabel}%
	\scr@startsection{section}{1}{\z@}{-3.5ex \@plus -1ex \@minus -0.2ex}{2.3ex \@plus 0.2ex}{\normalfont\bfseries}{#1}%
	\protected@write\@auxout{}{\string\@writefile{toc}{\string\contentsline {\reftype}{\string\numberline {\refnumber}#1}{\thepage}{\reflabel}}}%
}
\newcommand\articlesectionstar[1]{%
	\articlesectiondata{#1}%
	\scr@startsection{section}{1}{\z@}{-3.5ex \@plus -1ex \@minus -0.2ex}{2.3ex \@plus 0.2ex}{\normalfont\bfseries}*{#1}%
}
\newcommand\articlesubsectionnostar[1]{%
	\articlesubsectiondata{#1}%
	\pdfbookmark[2]{\reftext}{\reflabel}%
	\scr@startsection{subsection}{2}{\z@}{-3.5ex \@plus -1ex \@minus -0.2ex}{2.3ex \@plus 0.2ex}{\normalfont\bfseries}{#1}%
	\protected@write\@auxout{}{\string\@writefile{toc}{\string\contentsline {\reftype}{\string\numberline {\refnumber}#1}{\thepage}{\reflabel}}}%
}
\newcommand\articlesubsectionstar[1]{%
	\articlesubsectiondata{#1}%
	\scr@startsection{subsection}{2}{\z@}{-3.5ex \@plus -1ex \@minus -0.2ex}{2.3ex \@plus 0.2ex}{\normalfont\bfseries}*{#1}%
}
\newcommand\articlesection{\@ifstar{\stepcounter{othercounter}\articlesectionstar}{\articlesectionnostar}}
\newcommand\articlesubsection{\@ifstar{\stepcounter{othercounter}\articlesubsectionstar}{\articlesubsectionnostar}}
\renewcommand\chapter{\@ifstar{\stepcounter{othercounter}\oldchapter*}{\oldchapter}}
\renewcommand\section{\@ifstar{\stepcounter{othercounter}\oldsection*}{\oldsection}}
\renewcommand\subsection{\@ifstar{\stepcounter{othercounter}\oldsubsection*}{\oldsubsection}}
\renewcommand\subsubsection{\@ifstar{\stepcounter{othercounter}\oldsubsubsection*}{\oldsubsubsection}}
\newcommand\listof{}

\newcommand\articlefiguredata{%
	\renewcommand\listof{lof}%
	\renewcommand\reftype{figure}%
	\renewcommand\reflabel{figure.\arabic{chapter}.\arabic{figure}}%
	\renewcommand\refnumber{\Alph{otherchapter}.\arabic{figure}}%
	\renewcommand\refshortnumber{\arabic{figure}}%
}
\newcommand\articletabledata{%
	\renewcommand\listof{lot}%
	\renewcommand\reftype{table}%
	\renewcommand\reflabel{table.\arabic{chapter}.\arabic{table}}%
	\renewcommand\refnumber{\Alph{otherchapter}.\arabic{table}}%
	\renewcommand\refshortnumber{\arabic{table}}%
}
\renewenvironment{figure}{\articlefiguredata\begin{oldfigure}}{\end{oldfigure}} 
\renewenvironment{table}{\articletabledata\begin{oldtable}}{\end{oldtable}} 
\newenvironment{articlefigure}{\articlefiguredata\begin{figure}}{\end{figure}}
\newenvironment{articlefigure*}{\articlefiguredata\begin{figure*}}{\end{figure*}}
\newenvironment{articletable}{\articletabledata\begin{table}}{\end{table}}
\newenvironment{articletable*}{\articletabledata\begin{table*}}{\end{table*}}
\let\oldcaption\caption
\newcommand\Acaption[2][]{%
	\oldcaption[#1]{#2}%
	\renewcommand\reftext{#1}%
	\protected@write\@auxout{}{\string\@writefile{\listof}{\string\contentsline {\reftype}{\string\numberline {\refnumber}#1}{\thepage}{\reflabel}}}%
}
\renewcommand\caption[2][]{\ifarticlestyle\Acaption[#1]{#2}\else\oldcaption[#1]{#2}\fi}

\newcommand\botholdlabel[1]{\oldlabel{#1}\oldlabel{A#1}}
\newenvironment{articleequation}{\begin{equation}\renewcommand\label{\botholdlabel}}{\end{equation}} 
\newcommand\Aref[1]{\oldref{A#1}}
\newcommand\Alabel[1]{%
	\protected@write\@auxout{}{\string\newlabel{#1}{{\refnumber}{\thepage}{\reftext}{\reflabel}{}}}%
	\protected@write\@auxout{}{\string\newlabel{A#1}{{\refshortnumber}{\thepage}{\reftext}{\reflabel}{}}}%
}
\AtBeginDocument{%
	\let\oldref\ref%
	\let\oldlabel\label%
	\renewcommand\ref[1]{\ifarticlestyle\Aref{#1}\else\oldref{#1}\fi}%
	\renewcommand\label[1]{\ifarticlestyle\Alabel{#1}\else\oldlabel{#1}\fi}%
}
\makeatother

\newcommand\articlestyleheaderleft{%
	\ifx\insertarxiv\empty%
		\ifx\insertjournal\empty\linebreak\textnormal\insertdoi\else\linebreak\textnormal\insertjournal\fi%
	\else%
		\ifx\insertdoi\empty\linebreak\textnormal\insertjournal\else\textnormal\insertjournal\linebreak\textnormal\insertdoi\fi%
	\fi%
}
\newcommand\articlestyleheaderright{%
	\ifx\insertarxiv\empty%
		\ifx\insertjournal\empty\else\linebreak\textnormal\insertdoi\fi%
	\else%
		\linebreak\textnormal\insertarxiv%
	\fi%
}
\newcommand\articlestyleheadercenter{%
	\linebreak\textnormal\thechapter
}

\newcounter{articlepage}
\newcommand\articlepagemark{\arabic{articlepage}}

\newcommand\nocontentsline[3]{}
\let\oldaddcontentsline\addcontentsline
\newcommand\normalstyle{%
	\articlestylefalse%
	\KOMAoptions{fontsize=11pt}%
	\newgeometry{left=3cm,right=2.5cm,top=4cm,bottom=4cm}
	\setlength{\headheight}{26pt}
	\setlength{\headsep}{24pt}
	\setlength{\footskip}{30pt}
	\setlength{\TwoColumnWidth}{\textwidth}
	\setlength{\OneColumnWidth}{0.5\TwoColumnWidth-0.5\columnsep}
	\clearpairofpagestyles%
	\ihead{}%
	\chead{}%
	\ohead{\ifthispageodd{\textnormal\rightmark}{\textnormal\leftmark}}%
	\ifoot{}%
	\cfoot{}%
	\ofoot[\textnormal\pagemark]{\textnormal\pagemark}%
	\let\addcontentsline\oldaddcontentsline%
	\renewcommand{\thechapter}{\arabic{normalchapter}}%
	\renewcommand{\thesection}{\arabic{normalchapter}.\arabic{section}}%
	\renewcommand{\thesubsection}{\arabic{normalchapter}.\arabic{section}.\arabic{subsection}}%
	\renewcommand{\thesubsubsection}{\arabic{normalchapter}.\arabic{section}.\arabic{subsection}.\arabic{subsubsection}}%
	\renewcommand{\thefigure}{\arabic{normalchapter}.\arabic{figure}}%
	\renewcommand{\thetable}{\arabic{normalchapter}.\arabic{table}}%
	\renewcommand{\theequation}{\arabic{normalchapter}.\arabic{equation}}%
	\renewcommand\bibliographysection{\section*}%
	\renewcommand\bibcontentsline{\addcontentsline{toc}{section}{References}}
	\renewcommand\bibliographysectionstyle{}%
	\renewcommand\bibliographyitemsize{\normalsize}%
	\renewcommand\bibliographyitemseparation{%
		\setlength{\parskip}{0pt}%
		\setlength{\itemsep}{5pt plus 0.3ex}%
	}%
	\allowdisplaybreaks%
}
\newcommand\articlestyle{%
	\articlestyletrue%
	\KOMAoptions{fontsize=10pt}%
	\newgeometry{left=1.5cm,right=1.5cm,top=2.95cm,bottom=1.55cm}
	\setlength{\headheight}{24pt}
	\setlength{\headsep}{20pt}
	\setlength{\footskip}{1.2cm}
	\setlength{\TwoColumnWidth}{\textwidth}
	\setlength{\OneColumnWidth}{0.5\TwoColumnWidth-0.5\columnsep}
	\clearpairofpagestyles%
	\ihead{\ifthispageodd{\articlestyleheaderleft}{\articlestyleheaderright}}%
	\chead{\ifsinglepaper\else\articlestyleheadercenter\fi}%
	\ohead{\ifthispageodd{\articlestyleheaderright}{\articlestyleheaderleft}}%
	\ifoot{}%
	\cfoot{\ifsinglepaper\textnormal\pagemark\else\stepcounter{articlepage}\textnormal{\thechapter-\articlepagemark}\fi}%
	\ofoot{\ifsinglepaper\else\textnormal\pagemark\fi}%
	\let\addcontentsline\nocontentsline%
	\renewcommand{\thechapter}{\Alph{otherchapter}}%
	\renewcommand{\thesection}{\arabic{section}}%
	\renewcommand{\thesubsection}{\arabic{section}.\arabic{subsection}}%
	\renewcommand{\thesubsubsection}{\arabic{section}.\arabic{subsection}.\arabic{subsubsection}}%
	\renewcommand{\thefigure}{\arabic{figure}}%
	\renewcommand{\thetable}{\arabic{table}}%
	\renewcommand{\theequation}{\arabic{equation}}%
	\renewcommand\bibliographysection{\articlesection*}%
	\renewcommand\bibcontentsline{\oldaddcontentsline{toc}{section}{References}}
	\renewcommand\bibliographysectionstyle{\normalsize}%
	\renewcommand\bibliographyitemsize{\small}%
	\renewcommand\bibliographyitemseparation{%
		\setlength{\parskip}{0pt}%
		\setlength{\itemsep}{0pt plus 0.3ex}%
	}%
	\interdisplaylinepenalty=10000%
}
\normalstyle

\renewcommand{\hbar}{\mathchar'26\mkern-9mu \mathrm{h}}
\newcommand{\one}{\mathcal{I}}
\newcommand{\hamilton}{\mathcal{H}}
\newcommand{\overlap}{\mathcal{O}}

\newcommand{\hamiltoncoupling}{\tau}
\newcommand{\overlapcoupling}{\nu}
\newcommand{\green}{\mathcal{G}}

\newcommand{\eqalign}[1]{\begin{aligned}#1\end{aligned}}

\singlepapertrue
\begin{document}

\raggedbottom

\frontmatter
\clearpairofpagestyles
\ofoot[\textnormal\pagemark]{\textnormal\pagemark}
\KOMAoptions{headsepline=false}

\mainmatter
\KOMAoptions{headsepline=true}

\normalstyle
\articlestyle
\normalsize

\renewcommand{\one}{\mathcal{I}}
\renewcommand{\hamilton}{\mathcal{H}}
\renewcommand{\hamiltoncoupling}{\tau}
\renewcommand{\overlap}{\mathcal{S}}
\renewcommand{\overlapcoupling}{\sigma}
\renewcommand{\green}{\mathcal{G}}

\twocolumn 

\title{Electronic transport through defective semiconducting carbon nanotubes}{JPC1}

\addauthor{Fabian Teichert}{1,3,4}
\addauthor{Andreas Zienert}{2}
\addauthor{J\"org Schuster}{3,4}
\addauthor{Michael Schreiber}{1,4}

\addaddress{Institute of Physics, Chemnitz University of Technology, 09107 Chemnitz, Germany}
\addaddress{Center for Microtechnologies, Chemnitz University of Technology, 09107 Chemnitz, Germany}
\addaddress{Fraunhofer Institute for Electronic Nano Systems (ENAS), 09126 Chemnitz, Germany}
\addaddress{Dresden Center for Computational Materials Science (DCMS), TU Dresden, 01062 Dresden, Germany}

\email{fabian.teichert@physik.tu-chemnitz.de
}

\abstract{
We investigate the electronic transport properties of semiconducting ($m$,$n$) carbon nanotubes (CNTs) on the mesoscopic length scale with arbitrarily distributed realistic defects.
The study is done by performing quantum transport calculations based on recursive Green's function techniques and an underlying density-functional-based tight-binding model for the description of the electronic structure.
Zigzag CNTs as well as chiral CNTs of different diameter are considered.
Different defects are exemplarily represented by monovacancies and divacancies.
We show the energy-dependent transmission and the temperature-dependent conductance as a function of the number of defects.
In the limit of many defetcs, the transport is described by strong localization.
Corresponding localization lengths are calculated (energy dependent and temperature dependent) and systematically compared for a large number of CNTs.
It is shown, that a distinction by $(m-n)\,\mathrm{mod}\,3$ has to be drawn in order to classify CNTs with different bandgaps.
Besides this, the localization length for a given defect probability per unit cell depends linearly on the CNT diameter, but not on the CNT chirality.
Finally, elastic mean free paths in the diffusive regime are computed for the limit of few defects, yielding qualitatively same statements.
}

\addkeyword{carbon nanotube (CNT)}
\addkeyword{defect}
\addkeyword{density-functional-based tight binding (DFTB)}
\addkeyword{electronic transport}
\addkeyword{recursive Green's function formalism (RGF)}
\addkeyword{strong localization}
\addkeyword{elastic mean free path}

\journal{J. Phys. Commun. 2 (2018), 105012}{Journal of Physics Communications 2 (2018), 105012}{http://iopscience.iop.org/article/10.1088/2399-6528/aae4cb} 
\doi{10.1088/2399-6528/aae4cb} 
\arxiv{1806.02737}{cond-mat.mes-hall} 

\articletitle

\articlesection{Introduction}

Semiconducting carbon nanotubes (CNTs) are promising candidates for future microelectronic devices.
Their high aspect ratio, nanoscopic diameter, and stable structure makes them applicable as channel material in field effect transistors\cite{Nature.393.49, NanoLett.5.345, Carbon.91.370, ACSNano.8.8730}.
Because of their strain dependent bandgap, CNTs can be used for mechanical sensors~\cite{NanoLett.6.1449, JComputElectron.15.881}.
On the other hand, defects play an important role by influencing the tubes' electronic properties.
Under clean laboratory conditions it is possible to grow long and defect-free CNTs~\cite{PhysRevLett.87.106801}.
However, currently this is hardly possible in mass production processes at the wafer level.
Multiple productions steps like etching and plasma treatments favor the subsequent creation of defects~\cite{PhysRevB.63.245405, NatureMaterials.4.534, ComputMatSci.93.15, NanoLett.9.2285, JPhysDApplPhys.43.305402}, which have a large influence on the device performance~\cite{PhysRevLett.89.216801, ChemPhysLett.661.1}.
Consequently, it is of great interest to know the impact of defects on the electronic structure and transport properties.

In the following article, we describe the transmission and the conductance through semiconducting CNTs \cite{JPhysSocJap.74.777, JPhysSocJap.68.3146} with randomly positioned vacancy defects by performing quantum transport calculations based on a density-functional tight-binding (DFTB) model.
Previous experiments, in which defects were created by ion irradiation, indicate that for long CNTs the strong localization regime can be achieved, where the conductance decreases exponentially with the CNT length~\cite{NatureMaterials.4.534}.
This was also studied theoretically for different defect types~\cite{SolidStateCommun.149.874, PhysRevLett.100.176803, JPhysChemC.117.15266, PhysStatSolB.247.2962, NanoRes.3.288, NanoLett.9.940}, especially for vacancies~\cite{PhysRevLett.95.266801, JPhysCondMat.20.294214, JPhysCondMat.20.304211, JPhysCondMat.26.045303, JPhysChemC.116.1179}, and also for other materials like silicon nanowires~\cite{PhysRevLett.99.076803}.
First analytic derivations of White and Todorov showed that the localization exponent depends linearly on the tube diameter~\cite{Nature.393.240}.
Flores et al. verified this for the first time with quantum transport calculations for metallic CNTs~\cite{JPhysCondMat.20.304211}.
We confirmed this by a more comprehensive analysis~\cite{NJPhys.16.123026} and extended it to defect mixtures~\cite{ComputMatSci.138.49}.
An investigation of semiconducting zig-zag CNTs~\cite{JComputElectron.17.521} showed qualitatively similar results for the localization and the diffusive regime at fixed energy.
Previous work focused mostly on the diameter dependence.
The chirality dependence has not been investigated in detail so far.

In the present work, we study a large amount of CNTs covering a wide range of diameters and all possible chiral angles.
First, we show how the localization, the diffusive, and the transition regime can be described.
Afterwards, we calculate the corresponding localization lengths and the elastic mean free paths as functions of the tube diameter and the chiral angle.
Those results are determined from the energy-dependent transmissions, as usual.
We show that similar quantities can be extracted from the temperature-dependent conductance.
Finally, we derive analytic expressions for the above results valid for all semiconducting CNTs.

\begin{articlefigure*}
	\begin{minipage}{0.5\textwidth}(a)\end{minipage}%
	\begin{minipage}{0.5\textwidth}(b)\end{minipage}\\[-1.4em]
	\includegraphics{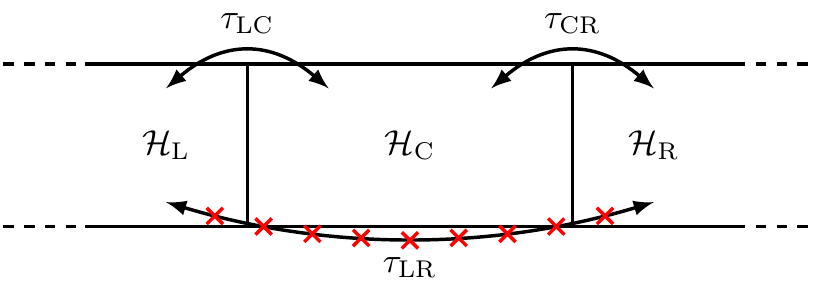}\hfill
	\includegraphics{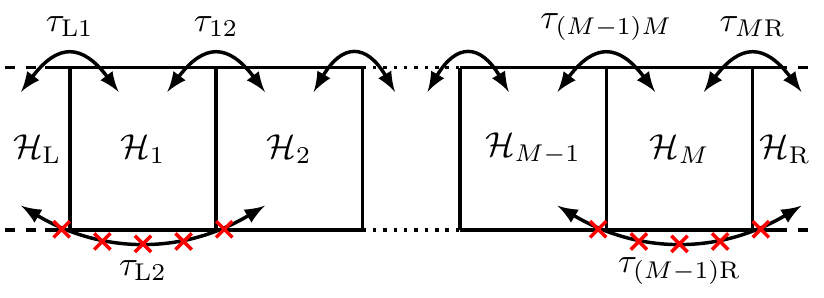}
	\caption[Scheme of a device system]{(color online) (a) Device configuration for the quasi-1D quantum transport theory~\cite{NJPhys.16.123026}. A finite central region C is connected to two half-infinite electrodes L and R. (b) Subdivision of C into $M$ blocks for the application of the RGF.}\label{JPC1:fig:Device}
\end{articlefigure*}

\articlesection{Theoretical framework}\label{JPC1:Theory}

The electronic transport is described by the equilibrium quantum transport theory for quasi-1D systems \cite{Datta2005}.
It is based on the device configuration shown in figure \ref{JPC1:fig:Device}(a), where a finite central region (C) is connected to two half-infinite electrodes (left L and right R).
This is an open system, where the electrodes act as reservoirs providing the electrons.
For this study, region C contains the part of the CNT with all the defects.
L and R are defect-free.
The Schr\"odinger equation (within an orthonormal basis) for the device reads
\begin{articleequation}
	\left(\begin{array}{ccc}
		\hamilton_\mathrm{L} & \hamiltoncoupling_\mathrm{LC} & 0 \\
		\hamiltoncoupling_\mathrm{CL} & \hamilton_\mathrm{C} & \hamiltoncoupling_\mathrm{CR} \\
		0 & \hamiltoncoupling_\mathrm{RC} & \hamilton_\mathrm{R} \\
	\end{array}\right) \mathit{\Psi} = E \mathit{\Psi} \quad .
\end{articleequation}%
$\hamilton_\mathrm{L/C/R}$ are the Hamiltonian matrices of the corresponding region.
$\hamiltoncoupling_\mathrm{LC/CL/RC/CR}$ are the coupling matrices between two of the regions.
Region C is chosen large enough, i.e. larger than the interaction distance, so that the direct coupling between the two electrodes $\hamiltoncoupling_\mathrm{LR/RL}$ can be neglected.
For a non-orthogonal basis with corresponding overlap matrices $\overlap_\mathrm{L/C/R}$ and overlap coupling matrices $\overlapcoupling_\mathrm{LC/CL/RC/CR}$ the Schr\"odinger equation and all subsequent equations can be obtained by substituting $\hamilton:=\hamilton-E\left(\overlap-\one\right)$ and $\hamiltoncoupling:=\hamiltoncoupling-E\overlapcoupling$.
$\one$ is the identity matrix of appropriate size.

The calculation of the transmission is based on the Green's function approach.
The Green's function of the central region is
\begin{articleequation}
	\green_\mathrm{C} = \left[ \left(E-\mathrm{i}\eta\right)\one - \hamilton_\mathrm{C} - \mathit{\Sigma}_\mathrm{L} - \mathit{\Sigma}_\mathrm{R} \right]^{-1} \quad .
\end{articleequation}%
$\eta$ is a small real number to shift the singularities at certain energies away from the real axis into the complex plane, which improves the convergence of the inversion.
We use $\eta=10^{-7}$ for the central region and $\eta=10^{-4}$ for the electrodes.
$\mathit{\Sigma}_\mathrm{L/R}$ are self-energy matrices for the left/right electrode leading to an energetic shift and broadening of the electronic states of C due to the coupling to L/R.
They can be calculated with
\begin{articleequation}
	\mathit{\Sigma}_\mathrm{L} = \hamiltoncoupling_\mathrm{CL}\green_\mathrm{L}\hamiltoncoupling_\mathrm{LC} \quad,\quad  \mathit{\Sigma}_\mathrm{R} = \hamiltoncoupling_\mathrm{CR}\green_\mathrm{R}\hamiltoncoupling_\mathrm{RC} \quad.
\end{articleequation}%
$\green_\mathrm{L/R}$ are the surface Green's functions of the electrodes.
They are calculated iteratively using the renormalization decimation algorithm (RDA)~\cite{JPhysFMetPhys.14.1205, JPhysFMetPhys.15.851}, which treats 1D bulk-like matrices very efficiently.
We use an improved version, which we derived for very long unit cells like in chiral CNTs~\cite{arxiv.1806.02039}.
With this, the transmission of electrons at a given energy $E$ is
\begin{articleequation}
	\mathcal{T} = \mathrm{Tr}\left[ \mathit{\Gamma}_\mathrm{R}\green_\mathrm{C}\mathit{\Gamma}_\mathrm{L}\green_\mathrm{C}^\dagger \right] \quad .
\end{articleequation}%
The broadening matrices $\mathit{\Gamma}_\mathrm{L/R} = \mathrm{i}\left(\mathit{\Sigma}_\mathrm{L/R}-\mathit{\Sigma}_\mathrm{L/R}^\dagger\right)$ describe the energetic broadening of the electronic states of C due to the coupling to L/R.

Because of the finite maximum interaction distance, the central region of the device can be divided into many very small blocks.
This is shown in figure \ref{JPC1:fig:Device}(b).
The corresponding Hamiltonian $\hamilton_\mathrm{C}$ is blockwise tridiagonal.
One of these blocks must not necessarily contain multiples of whole unit cells.
They can in general be arbitrary.
The larger chiral and defective CNT cells are further subdivided into as many blocks as possible which are not shorter than the interaction cutoff distance.
With this, the transmission can be calculated recursively by using the recursive Green's function formalism (RGF)~\cite{JPhysCSolidStatePhys.14.235} with
\begin{articleequation}
	\mathcal{T} = \mathrm{Tr}\left[ \mathit{\Gamma}_\mathrm{R}'\green_{M1}\mathit{\Gamma}_\mathrm{L}'\green_{M1}^\dagger \right] \quad .
\end{articleequation}%
$\mathit{\Gamma}_\mathrm{L/R}'$ are the upper left and lower right blocks of $\mathit{\Gamma}_\mathrm{L/R}$.
$\green_{M1}$ is the lower left block of the total Green's matrix $\green_\mathrm{C}$.
Its dimension is smaller by a factor $M$.
$\green_{M1}$ can be computed with the RGF without calculating the other blocks of $\green_\mathrm{C}$, which saves much time concerning the matrix inversion.
For the case of few randomly distributed defects, the RGF can be improved by using RDA steps to treat the periodic parts between the defects~\cite{JComputPhys.334.607}.

\stepcounter{mpfootnote}
\begin{articlefigure*}[t]
	\includegraphics[width=\textwidth]{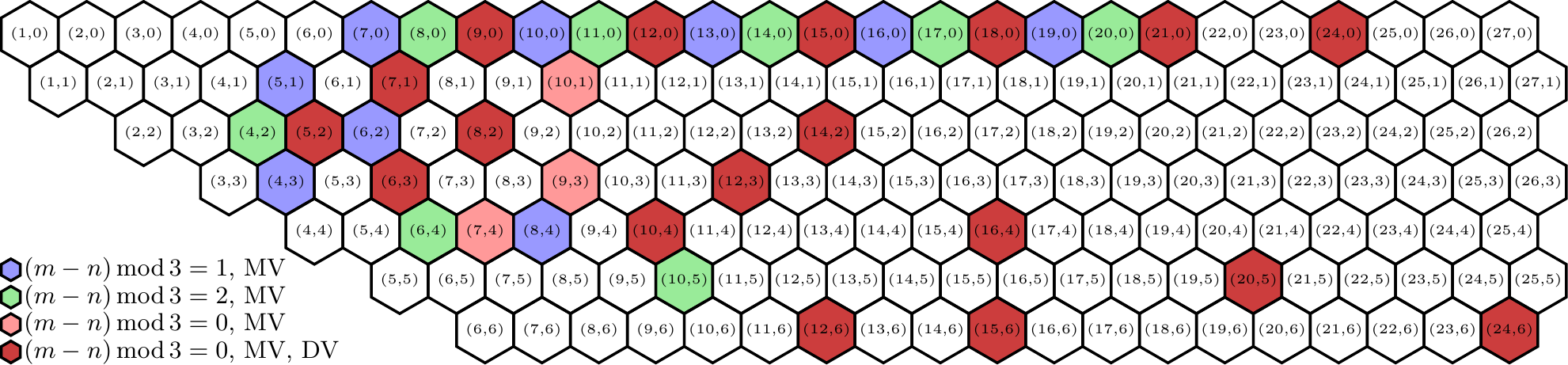}\\[1em]
	\includegraphics{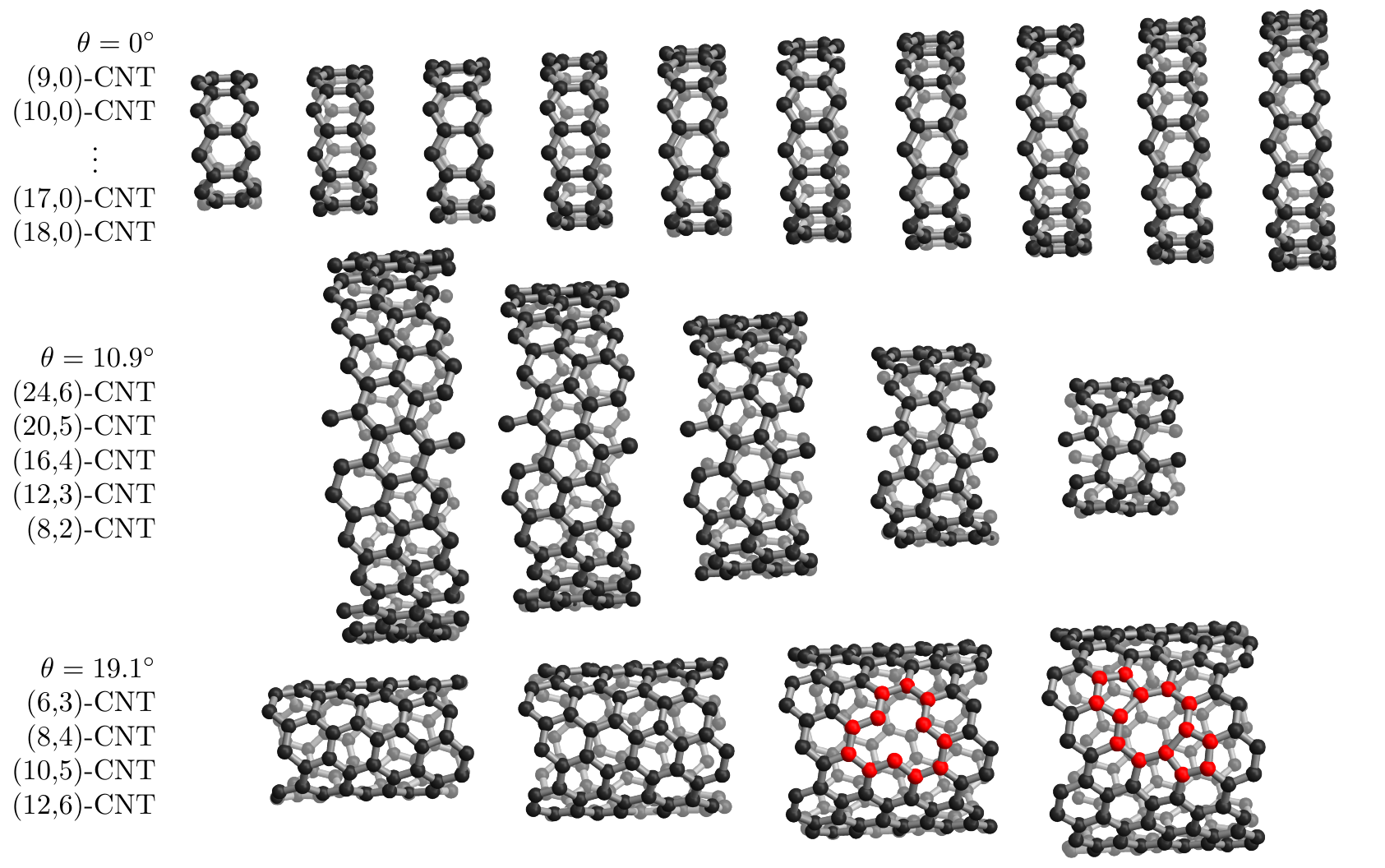}
	\caption[Periodic table of CNTs and geometric structures of the CNT cells]{(color online) (top) Periodic table of CNTs where the studied ones are marked by color. Different colors denote different subsets $(m-n)\,\mathrm{mod}\,3$. For the CNTs with light color, MV defects are considered, for the ones with dark color, both MV and DV defects are considered separately\footnotemark. (bottom) Unit cells of some exemplary CNTs with three different chiral angles $\theta$. The 2 lower right CNTs show the structure of the defects. The left one is the MV defect, the right one the DV defect. Atoms which surround the defect are colored in red.}\label{JPC1:fig:CNT}
\end{articlefigure*}

To calculate the above mentioned quantities we use a DFTB model~\cite{PhysRevB.51.12947, IntJQuantumChem.58.185} to describe the electronic structure and to calculate $\hamilton$, $\hamiltoncoupling$, $\overlap$, and $\overlapcoupling$ instantaneously.
This gives the speed of TB calculations but with DFT accuracy and thus makes quantitative statements possible.
For this work, the parameter set 3ob~\cite{JChemTheoryComput.9.338, DFTB} is used.
It is a non-orthogonal sp$^3$ basis set for organic molecules, which is especially suitable for aromatic carbon rings as the parameters have been obtained from DFT calculations for, e.g., benzene.
As the parameters are rapidly decreasing with increasing atom distance, a distinct interaction cutoff distance can be used.
We utilize twice the graphene carbon-carbon distance $a_\mathrm{CC}=1.42\,\text{\AA}$, which leads to a third-nearest-neighbor TB Hamiltonian with distance-dependent hopping parameters.
The smallest possible block contains exactly one unit cell of a zigzag CNT.

The calculations within this work are done in the low bias limit.
Furthermore, phonons are neglected, which means that the results are limited to CNT lengths smaller than the coherence length.
This is justified because optical phonons have too high energies to be excited thermally at room temperature and acoustic phonons can have coherence lengths up to a few $\upmu$m~\cite{PhysRevLett.94.086802,PhysRevLett.104.116801}.

\articlesection{Modeling details}

\begin{articlefigure*}
	\includegraphics{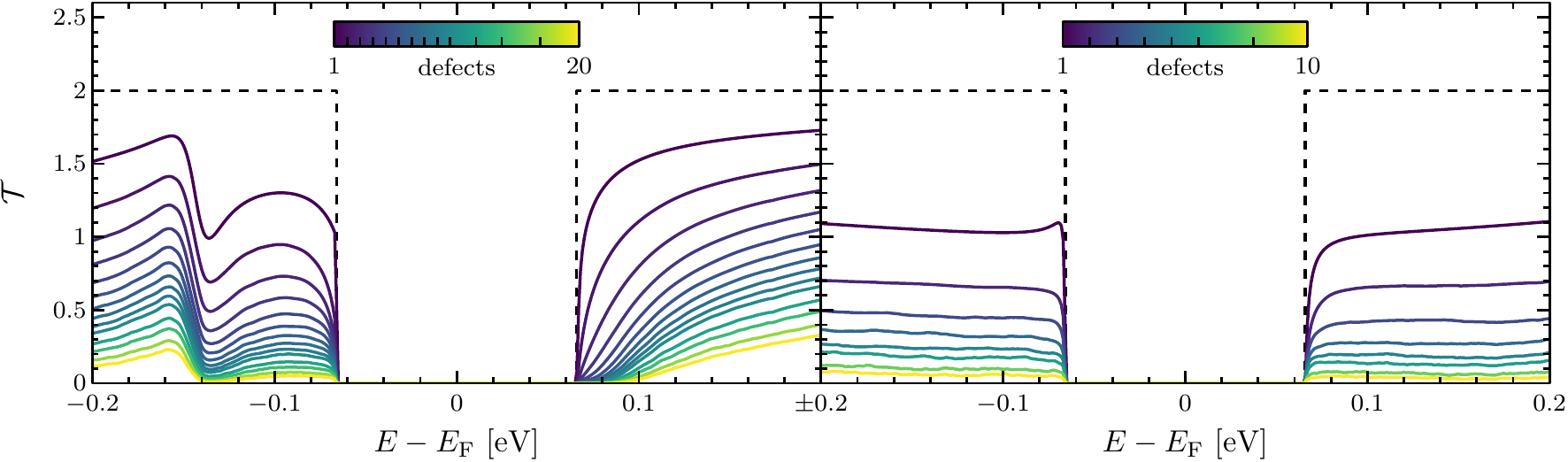}
	\caption[Transmission spectrum of the (9,0)-CNT with MV and DV defects]{(color online) Transmission spectrum of the (9,0)-CNT with 1, 2, 3,\ldots, 10, 12, 14, 17, 20 MV defects (left, from top to bottom) and with 1, 2, 3, 4, 5, 6, 8, 10 DV defects (right, from top to bottom).}\label{JPC1:fig:T(E)}
\end{articlefigure*}

\footnotetext{Mixtures of defects were studied previously~\cite{ComputMatSci.138.49}}

To describe the structural influence we consider\linebreak ($m$,$n$)-CNTs of different tube diameters\linebreak $d=\sqrt{3}a_\mathrm{CC}\sqrt{m^2+n^2+mn}/\pi$, different chiral angles\linebreak $\theta=\arctan[\sqrt{3}n/(2m+n)]$, and different subsets concerning $(m-n)\,\mathrm{mod}\,3$ (this distinction is justified later).\linebreak
In total 38 CNTs are investigated, as highlighted in\linebreak the periodic table of CNTs in figure \ref{JPC1:fig:CNT}, covering\linebreak $4.2\,\text{\AA}<d<22\,\text{\AA}$ and 11 different chiral angles\linebreak $0^\circ\leq\theta\leq 30^\circ$.
Some of the structures are shown in figure~\ref{JPC1:fig:CNT}.

The exemplarily studied defects, namely monovacancies (MV) and divacancies (DV), are both depicted in the lower right of figure \ref{JPC1:fig:CNT}.
The MV is created by removing one atom.
This defect has a small extension and fits within one unit cell.
The DV is created removing two neighbored atoms.
It is one of the common defects created, e.g., by ion bombardment~\cite{PhysRevB.63.245405}.
For this defect, three different orientations exist concerning the three different chiral carbon-carbon bond directions.
For all defects model structures are obtained by a geometry optimization of the directly surrounding atoms (red in figure \ref{JPC1:fig:CNT}) and the directly adjacent hexagons.
The DV defect extension is larger than the one of the MV.
Thus, for the CNT types with short unit cells (i.e. $\theta=0^\circ$ and $\theta=10.9^\circ$) the DV cell is two or three times larger in order to contain the whole defect.
The influence of the defect cell size and resulting long-range deformations on the electron transport was already investigated~\cite{arxiv.1705.01753}.

The geometry optimization is performed using density functional theory within the implementation of Atomistix ToolKit~\cite{ATK.12.8.2, PhysRevB.65.165401}.
For this, the local density approximation of Perdew and Zunger~\cite{PhysRevB.23.5048}, norm-conserving Troullier-Martins pseudopotentials~\cite{PhysRevB.43.1993}, and a SIESTA type double zeta plus double polarization basis set~\cite{JPhysCondMat.14.2745} are used.

The defects are randomly distributed within the CNT.
For this, the length of the device central region is fixed to $852\,\mathrm{nm}$ for the zigzag CNTs (i.e. 2000 cells) and to similar lengths that match multiples of the unit cell length for the chiral CNTs.
Within this region, $N$ defects of one type (MV or DV) are positioned at random lateral positions, angular positions, and orientations.
An ensemble of 1000 such configurations is created to describe the transmission in the sense of an ensemble average and as a function of the number of defects $N$.
The electrodes are defect-free CNTs of the same type as in the central region.
With this we omit the influence of contact effects and describe the pure defect influence.

\articlesection{Results and discussion}

\articlesubsection{Transmission and transport regimes}\label{JPC1:sec:T}

The transmission function $\mathcal{T}(E)$ for one single CNT with randomly distributed defects depends strongly on their exact positions and alignments.
It is intuitive that the transmission should lower with increasing number of defects.
In fact the defect states introduce resonances into the system due to quantum interference.
The more defects the system has, the more and the sharper the resonances are, leading to an accumulation of random peaks, preventing a reasonable analysis of the results~\cite{PhysRevB.27.831}.
In the following we always analyze ensemble averages, which leads to smooth curves.
We omit the averaging symbol for simplicity.

\begin{articlefigure}[!b]
	\includegraphics{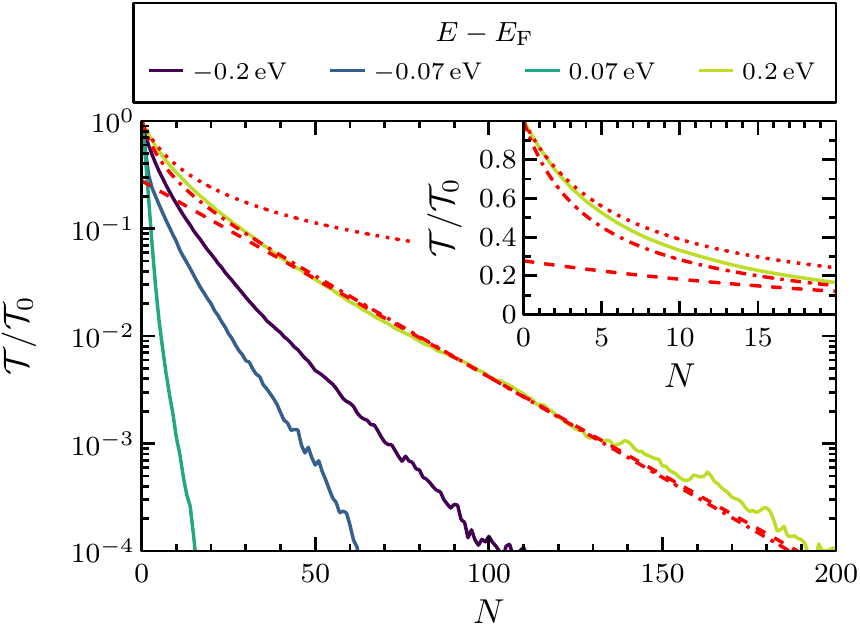}
	\caption[Transmission of the (9,0)-CNT as a function of the number of MV defects]{(color online) Transmission at a fixed energy of the (9,0)-CNT as a function of the number of MV defects. The energies $E=E_\mathrm{F}\pm 0.07$ eV lie near the valence/conduction band edge, $E=E_\mathrm{F}\pm 0.2$ eV lie far away from the band edges. The red dashed line is a regression in the localization regime (\ref{JPC1:eqn:Tloc2}), the red dotted line in the diffusive regime (\ref{JPC1:eqn:Tdif}), and the red dash-dotted line in the combined regime (\ref{JPC1:eqn:Tlocdif}) including the transition region. The inset shows the transmission for a few defects.}\label{JPC1:fig:T(N)}
\end{articlefigure}

Figure \ref{JPC1:fig:T(E)} shows the (average) transmission function for the (9,0)-CNT with 1 up to 20 defects.
A systematically decreasing transmission can be seen as well as a defect-induced resonance for the MV around $140\,\mathrm{meV}$ below the Fermi energy, which leads to a dip.
Figure \ref{JPC1:fig:T(N)} shows the dependence on the number of defects $N$, normalized to the bulk transmission $\mathcal{T}_0$, which equals the number of conductance channels.
Transmissions calculated at four different energies are depicted: the valence band edge, the conduction band edge and two energies within the bands.
A clear exponential decrease can be seen for sufficiently large values of $N$.
This can be explained by the strong localization regime, where the electronic states are exponentially localized due to destructive interference with a characteristic localization length $\ell^\mathrm{loc}$.
For Anderson-like disorder it was shown that 1D systems are always in the strong localization regime~\cite{AdvPhys.10.107, PhysRevLett.47.1546, PhysRevLett.42.673}.
As a consequence the transport is exponentially suppressed.
For a fixed defect probability and varying system length $L$ the transport in the limit of large $L$ resp. small $\mathcal{T}$ can be described by
\begin{articleequation}
	\mathcal{T} \propto \mathrm{e}^{-L/\ell^\mathrm{loc}} \quad .\label{JPC1:eqn:Tloc}
\end{articleequation}%
The same exponential dependence holds for fixed length and varying defect probability or -- in our case -- number of defects, as shown in~\cite{NJPhys.16.123026, JComputElectron.17.521}:
\begin{articleequation}
	\mathcal{T} \propto \mathrm{e}^{-N/N^\mathrm{loc}} \quad .\label{JPC1:eqn:Tloc2}
\end{articleequation}%
The localization exponent $N^\mathrm{loc}$ is a material constant, which describes the number of additional defects needed to lower the transmission by a factor of $\mathrm{e}$.
The fluctuations visible in figure \ref{JPC1:fig:T(N)} for $\mathcal{T}/\mathcal{T}_0<10^{-3}$ are merely caused by the ensemble not being large enough for sampling a very high number of defects.

The systematic deviations from the exponential behavior for few defects at high transmission $\mathcal{T}/\mathcal{T}_0>10^{-1}$ are caused by an increasing localization length $\ell^\mathrm{loc}$.
For decreasing $N$ and fixed system length $L$ of the central region, the defect probability per cell decreases.
Thus, $\ell^\mathrm{loc}$ becomes of the same order or larger than $L$ at small $N$ and the system is no more in the strong localization regime.
This range of high transmission can be described by the diffusive regime, where only elastic scattering occurs, but without the long-range destructive interference effects limiting the transport.
In the diffusive regime the transport can be described by a resistance, which increases linear with the system length for fixed defect density~\cite{RevModPhys.69.731}.
For our case, where we fix the system length, we get for the transmission
\begin{articleequation}
	\frac{\mathcal{T}}{\mathcal{T}_0} = \left(1+\frac{L}{\ell^\mathrm{mfp}}\right)^{-1} = \left(1+\frac{N}{N^\mathrm{mfp}}\right)^{-1} \quad .\label{JPC1:eqn:Tdif}
\end{articleequation}%
$\ell^\mathrm{mfp}$ is the elastic mean free path, while $N^\mathrm{mfp}$ is the dimensionless elastic mean free path.

Both the diffusive and the localized regime, can be described phenomenologically by solving the steady-state diffusion equation with an additional sink term, which traps the diffusing electrons and hinders them to pass lengths larger than $\ell^\mathrm{loc}$: $0=n''(x)-n(x)/\ell^\mathrm{loc}$, where $x$ is the position along the 1D system and $n$ is the 1D electron density.
The local current density can be calculated using $j(x)=-Dn'(x)$ with the diffusion constant $D=\ell^\mathrm{mfp}v/2$ and the average particle velocity $v$.
The trapped fraction of the current is $j_\mathrm{trap}=D/(\ell^\mathrm{loc})^2\cdot\int_0^Ln(x)\mathrm{d}x=j(0)-j(L)$.
With the appropriate boundary conditions $n(L)v=j_\mathrm{out}$ and $n(0)v=j_\mathrm{in}+j_\mathrm{ref}=2j_\mathrm{in}-j_\mathrm{out}-j_\mathrm{trap}$ for the ingoing/outgoing/reflected current density $j_\mathrm{in/out/ref}$ of a device configuration and $\mathcal{T}=j_\mathrm{out}/j_\mathrm{in}$, the transmission can be obtained~\cite{PhysRevB.54.5801}:
\begin{articleequation}
	\eqalign{
		\frac{\mathcal{T}}{\mathcal{T}_0} &= \left( \cosh \frac{L}{\ell^\mathrm{loc}} + \frac{\ell^\mathrm{loc}}{\ell^\mathrm{mfp}} \sinh \frac{L}{\ell^\mathrm{loc}} \right)^{-1} \\
		&= \left( \cosh \frac{N}{N^\mathrm{loc}} + \frac{N^\mathrm{loc}}{N^\mathrm{mfp}} \sinh \frac{N}{N^\mathrm{loc}} \right)^{-1} \quad .\label{JPC1:eqn:Tlocdif}
	}
\end{articleequation}%
In the limit $L \gg \ell^\mathrm{loc}$ this simplifies to (\ref{JPC1:eqn:Tloc}), in the limit $L \ll \ell^\mathrm{loc}$ to (\ref{JPC1:eqn:Tdif}).
The strong localization regime is approximately valid for $N>N^\mathrm{loc}$, the diffusive regime for $N<N^\mathrm{loc}/2$\footnotemark.

\begin{articlefigure*}
	\includegraphics{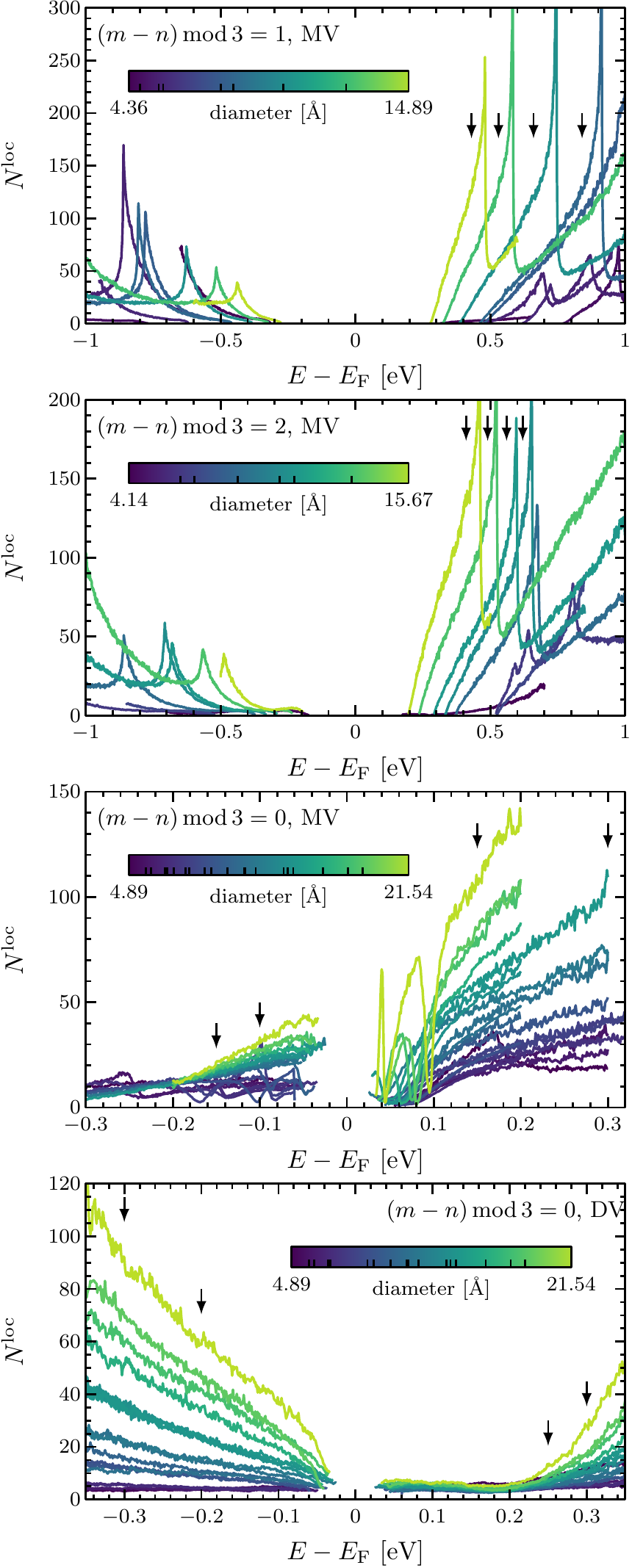}\hfill
	\includegraphics{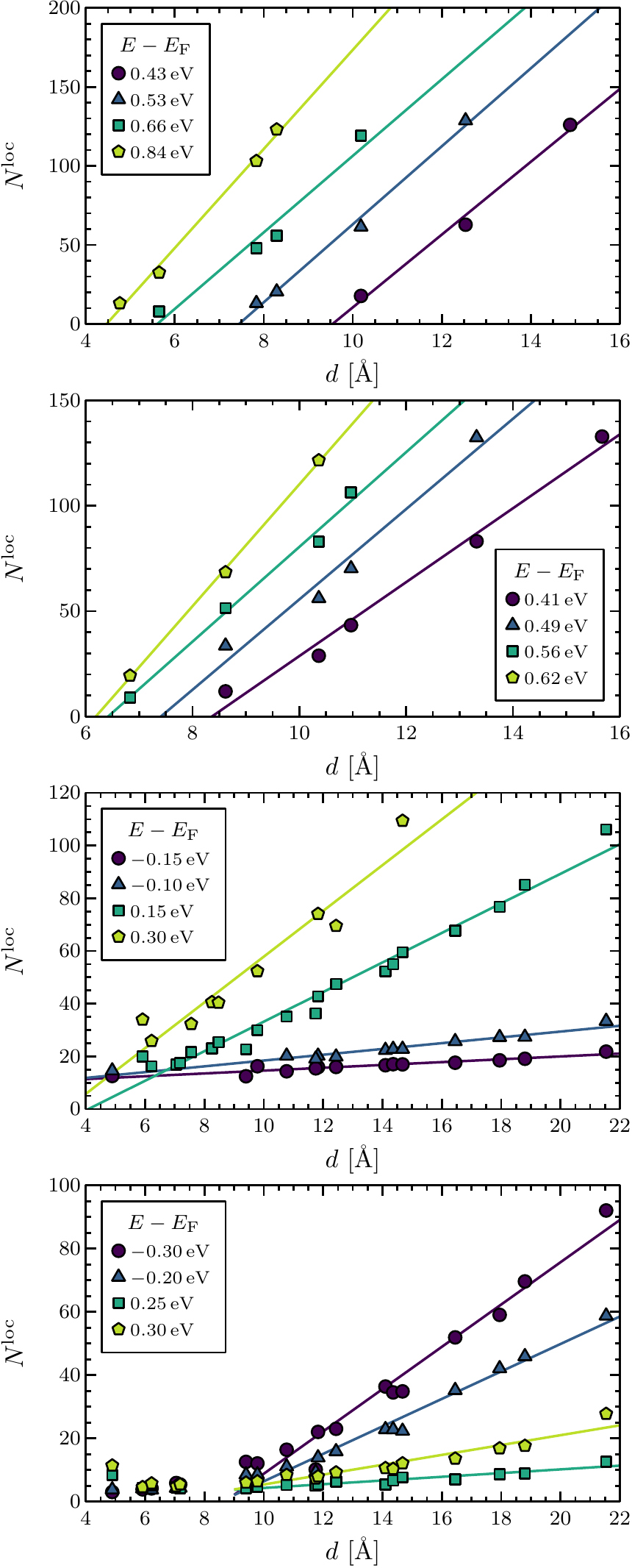}
	\caption[Localization exponent as a function of energy and diameter]{(color online) Left column: Localization exponent as a function of energy for the CNTs highlighted in figure \ref{JPC1:fig:CNT}. Different diameters are denoted by color. Right column: Diameter dependence of the localization exponent at fixed energies, which are marked by arrows in the left column. Different energies are denoted by color. The straight lines are linear regressions. Different rows show different subsets $(m-n)\,\mathrm{mod}\,3$ and different defects (MV or DV).}\label{JPC1:fig:loc(E)}
\end{articlefigure*}

An example for the transmission $\mathcal{T}(N)$ at $E=E_\mathrm{F}+0.2\,\mathrm{eV}$ is shown in figure \ref{JPC1:fig:T(N)}.
Equation (\ref{JPC1:eqn:Tlocdif}) (red dash-dotted line) describes the general shape of the data very well.
A corresponding regression gives $N^\mathrm{loc}=23$ and\linebreak $N^\mathrm{mfp}=4.2$.
Both regimes, localization (\ref{JPC1:eqn:Tloc2}) and diffusion (\ref{JPC1:eqn:Tdif}), are also depicted (red dashed and dotted lines) and fit well in the appropriate limit\footnote{A lower limit of the strong localization regime can be estimated comparing (\ref{JPC1:eqn:Tloc2}) and (\ref{JPC1:eqn:Tlocdif}).
	The relative error is smaller than $\epsilon$ if $N>N^\mathrm{loc}\ln((1-\frac{N^\mathrm{mfp}}{N^\mathrm{loc}})/\sqrt{\epsilon})$.
	For $\epsilon=10\%$ and $N^\mathrm{loc}/N^\mathrm{mfp}\approx 5$ this is roughly the case if $N>N^\mathrm{loc}$.
	The same can be done for the diffusive regime by comparing (\ref{JPC1:eqn:Tdif}) and (\ref{JPC1:eqn:Tlocdif}).
	The relative error is smaller than $\epsilon$ if $N<N^\mathrm{loc}\sqrt{6\epsilon}-N^\mathrm{mfp}$.
	For $\epsilon=10\%$ and $N^\mathrm{loc}/N^\mathrm{mfp}\approx 5$ this is roughly the case if $N<N^\mathrm{loc}/2$.}.
The regressions yield $N^\mathrm{loc}=24$ and $N^\mathrm{mfp}=6.4$.
The discrepancy concerning $N^\mathrm{mfp}$ comes from the fact, that (\ref{JPC1:eqn:Tlocdif}) underestimates the transition regime and gives a systematically overestimated derivative at $N=0$.
This can be seen in the inset of figure \ref{JPC1:fig:T(N)}.
Because of this we determine $N^\mathrm{loc}$ and $N^\mathrm{mfp}$ separately using (\ref{JPC1:eqn:Tloc2}) and (\ref{JPC1:eqn:Tdif}).
In detail, we obtain $N^\mathrm{loc}$ for each energy with a linear regression of the logarithmic data in the range $10^{-3}<\mathcal{T}/\mathcal{T}_0<10^{-1}$.
$N^\mathrm{mfp}$ is the slope of the linear dependence $\mathcal{T}^{-1}(N)$ and is calculated in the limit $N\rightarrow 0$.

\articlesubsection{Energy dependent localization exponent and elastic mean free path}\label{JPC1:sec:loc(E)}

Like in the preceding example the localization exponent is calculated for each energy and for all the different CNTs of figure \ref{JPC1:fig:CNT}.
The results are shown in the left column of figure \ref{JPC1:fig:loc(E)} for the MV (first three rows) and the DV (last row).
Here, a distinction between different subsets concerning $(m-n)\,\mathrm{mod}\,3$ must be made to qualify and quantify further dependencies.
The CNTs with\linebreak $(m-n)\,\mathrm{mod}\,3=0$ are the semi-metallic ones with very small bandgaps in the range $(50\ldots 130)\,\mathrm{meV}$.
The CNTs with $(m-n)\,\mathrm{mod}\,3=1$ and $(m-n)\,\mathrm{mod}\,3=2$ are the true semiconducting CNTs, the former ones with bandgaps in the range $(550\ldots 1540)\,\mathrm{meV}$ and the latter ones with bandgaps in the range $(340\ldots 1040)\,\mathrm{meV}$.

For each subset clear trends can be seen.
The qualitative shape of $N^\mathrm{loc}(E)$ is mostly the same: small values at the valence/conduction band edge and increasing values with decreasing/increasing energy up to the next band edge with singularity-like shapes there.
Furthermore, the curves follow a clear trend with varying diameter for most of the energies.
There are two exceptions: (a)~near the band edges, where $N^\mathrm{loc}$ diverges, and (b) near defect states, which lead to peaks/dips in the transmission spectrum and in $N^\mathrm{loc}(E)$.
The latter is the case for $(m-n)\,\mathrm{mod}\,3=0$: for the MV at energies below $0.1\,\mathrm{eV}$ and for the DV at energies between $0.0\,\mathrm{eV}$ and $0.2\,\mathrm{eV}$.
The right column of figure \ref{JPC1:fig:loc(E)} depicts the diameter dependence of $N^\mathrm{loc}$ at selected energies not fulfilling (a) or (b).
For all the cases a very clear linear dependence can be seen with only small deviations from the linear regressions (solid lines).
This is in good agreement with~\cite{JComputElectron.17.521}{\renewcommand{\thefootnote}{$\ast$}\footnote{The much larger deviations in \cite{JComputElectron.17.521} can be explained by the different bandgaps and by our subset distinction.}}.
The DV and $(m-n)\,\mathrm{mod}\,3=0$ case shows that there is a lower limit, here approximately $d=9\,\mathrm{nm}$ and\linebreak $N^\mathrm{loc}=6$.
These deviations for smaller tube diameters can be explained by strong curvature effects and the resulting changes of the $\pi$-bonds.
Also the fact that the defect occupies a large part of the CNT circumference can result in additionally distorted structures and changed transport properties.
Furthermore, it is important to mention that many different chiralities are included, especially for the $(m-n)\,\mathrm{mod}\,3=0$ case.
As the diameter dependence matches very well, a chirality dependence can be excluded.
In conclusion, the slope of $N^\mathrm{loc}(d)$ only depends on the subset, energy, and defect type, but not on the chirality.
The prefactor in (\ref{JPC1:eqn:Tloc2}) is between 0.4 and 1.2\footnote{	It strongly fluctuates as small changes in the localization exponent cause large changes in the prefactor. It depends less on energy, tube diameter, and chirality.}.

$N^\mathrm{mfp}$ is calculated in dependence on the energy in the same way as done before for $N^\mathrm{loc}$.
The result is shown exemplarily for the (16,0)-CNT in figure \ref{JPC1:fig:loc,dif(E)} in comparison to $N^\mathrm{loc}$.
It can be seen that $N^\mathrm{mfp}$ follows the general trend of $N^\mathrm{loc}$.
This is in agreement with former general studies, in which relations between the localization length and the elastic mean free path were derived~\cite{Nature.393.240, RevModPhys.69.731}.
Furthermore, a linear diameter dependence at fixed energy holds for $N^\mathrm{mfp}$, too.
We checked this for all studied CNTs with the same general result.

\begin{articlefigure}
	\includegraphics{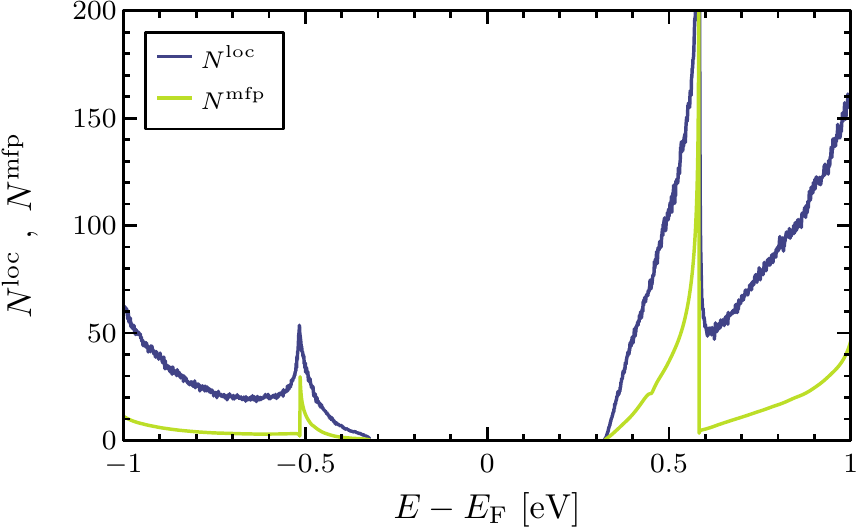}
	\caption[Localization exponent and elastic mean free path as a function of energy]{(color online) Localization exponent $N^\mathrm{loc}$ and normalized mean free path $N^\mathrm{mfp}$ as a function of energy for the (16,0)-CNT with MV defects.}\label{JPC1:fig:loc,dif(E)}
\end{articlefigure}

\begin{articletable}
	\begin{tabular}{l|c|c}
		& MV & DV\\
		\hline
		$(m-n)\,\mathrm{mod}\,3=0$ & \phantom{4}$-6\%$ \ldots $39\%$ & $-5\%$ \ldots $30\%$ \\
		$(m-n)\,\mathrm{mod}\,3=1$ & $-45\%$ \ldots $36\%$ & --- \\
		$(m-n)\,\mathrm{mod}\,3=2$ & $-46\%$ \ldots $34\%$ & --- \\
	\end{tabular}
	\caption[Deviations of $2/(1+N^\mathrm{loc}/N^\mathrm{mfp})$ compared to the regression prefactor of (\ref{JPC1:eqn:Tloc})]{Deviations of $2/(1+N^\mathrm{loc}/N^\mathrm{mfp})$ compared to the regression prefactor of the localized regime (\ref{JPC1:eqn:Tloc}).}
	\label{JPC1:tab:deviations}
\end{articletable}

A way for comparing $N^\mathrm{mfp}$ and $N^\mathrm{loc}$ is (\ref{JPC1:eqn:Tlocdif}) in the limit $N\gg N^\mathrm{loc}$.
This gives the prefactor $2/(1+N^\mathrm{loc}/N^\mathrm{mfp})$.
With this, the three regression parameters $N^\mathrm{mfp}$ from (\ref{JPC1:eqn:Tdif}), $N^\mathrm{loc}$ from (\ref{JPC1:eqn:Tloc2}), and the corresponding prefactor from (\ref{JPC1:eqn:Tloc2}) can be compared.
For all studied CNTs and energies, except near the band edges, where the regressions are less trustable, we get a good agreement.
The maximum deviations of $2/(1+N^\mathrm{loc}/N^\mathrm{mfp})$ compared to the regression prefactor are in the range of $\pm$40\%.
Details are listed in table \ref{JPC1:tab:deviations}.
This comparison shows that the determination of the three parameters is consistent, (\ref{JPC1:eqn:Tlocdif}) can be used to describe both regimes, and the prefactor of the localization regime can be estimated using $N^\mathrm{loc}$ and $N^\mathrm{mfp}$ with acceptable errors.

The linear dependencies $N^\mathrm{loc}(d)$ and $N^\mathrm{mfp}(d)$ can be used to predict the transmission of large diameter CNTs of any chirality.
But as $\mathcal{T}$ is energy-dependent this is not very practicable.

\articlesubsection{Conductance, effective localization exponent and effective elastic mean free path}

\begin{articlefigure}
	\includegraphics{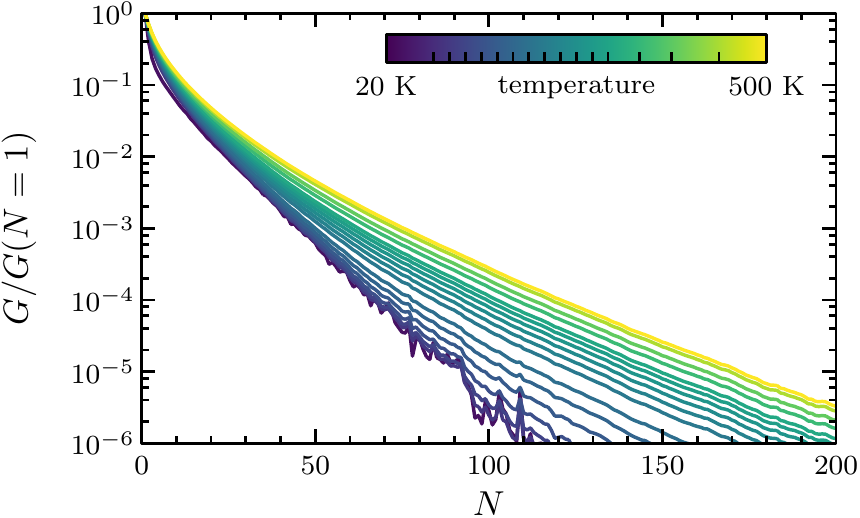}
	\caption[Conductance of the (9,0)-CNT as a function of the number of MV defects]{(color online) Conductance of the (9,0)-CNT as a function of the number of MV defects for different temperatures (denoted by color).}\label{JPC1:fig:G(N)}
\end{articlefigure}

In the mesoscopic range, the zero-bias conductance of an arbitrary scattering region between two reservoirs can be calculated using the Landauer-B\"uttiker formula~\cite{PhysRevB.31.6207}
\begin{articleequation}
	G = -\mathrm{G}_0\int\limits_{-\infty}^\infty\mathcal{T}(E)\frac{\mathrm{d}f(E)}{\mathrm{d}E}\mathrm{d}E \quad .\label{JPC1:eqn:LBF}
\end{articleequation}%
$\mathrm{G}_0=2\mathrm{e}^2/\mathrm{h}$ is the conductance quantum and $f(E)$ is the Fermi distribution, whereby the effect of temperature is included.

\begin{articlefigure*}[!t]
	\includegraphics{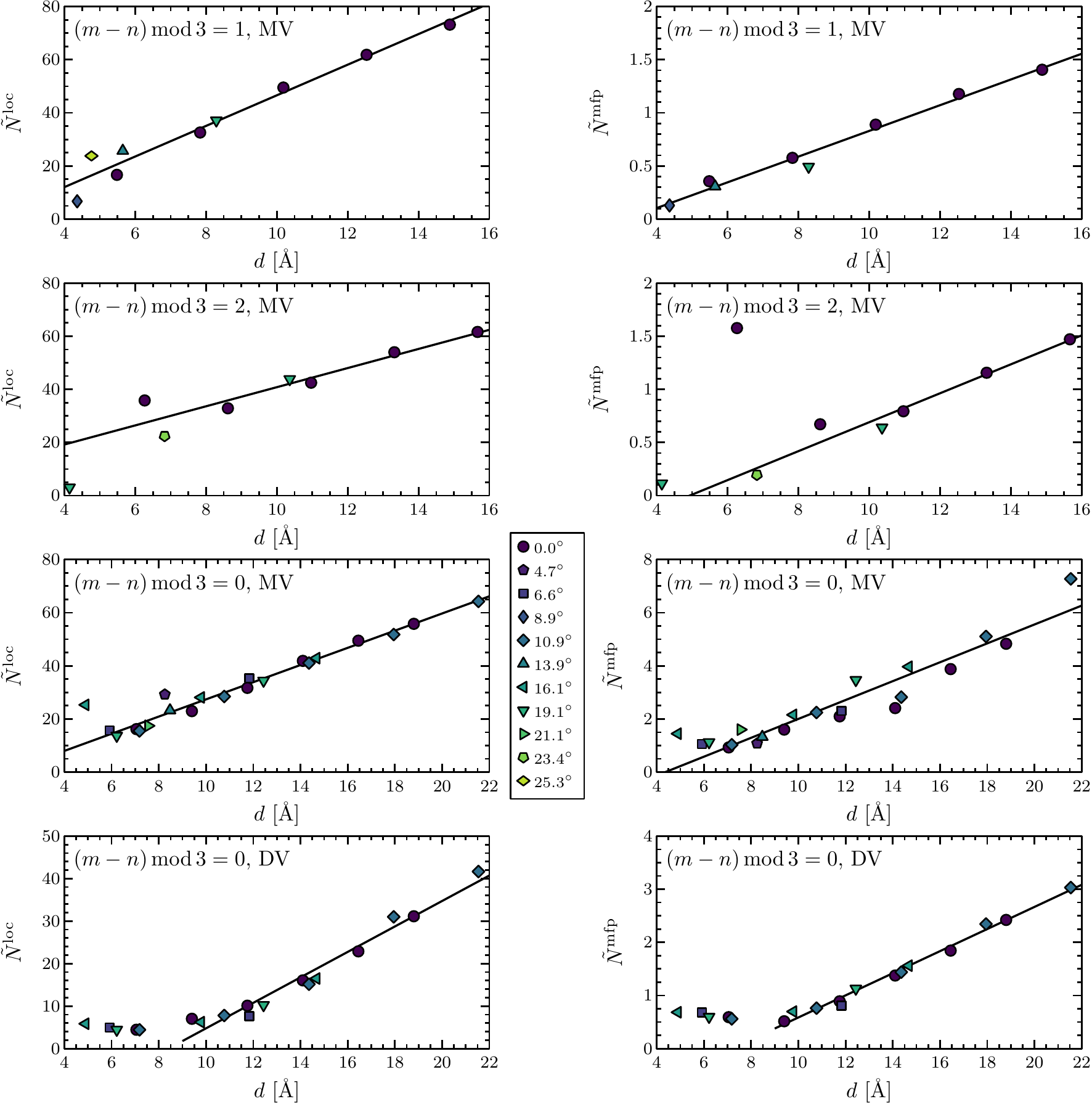}
	\caption[Localization exponent and elastic mean free path as a function of diameter]{(color online) Diameter dependence of (left column) the effective localization exponent and (right column) the effective elastic mean free path at $T=300\,\mathrm{K}$. Different chiral angles are denoted by color. The straight black lines are linear regressions. Different rows show different subsets $(m-n)\,\mathrm{mod}\,3$ and different defects (MV or DV).}\label{JPC1:fig:loc,dif(d)}
\end{articlefigure*}

An example for the conductance as a function of the number of defects is shown in figure \ref{JPC1:fig:G(N)} for the (9,0)-CNT for different temperatures.
For better comparison, all the data are normalized to the conductance of the (9,0)-CNT with one defect $G(N=1)$.
A similar picture as shown in figure \ref{JPC1:fig:T(N)} with a different effective localization exponent $\tilde{N}^\mathrm{loc}$ and elastic mean free path $\tilde{N}^\mathrm{mfp}$ can be seen.
In contrast to the transmission in the strong localization regime, the dependence on the defect number is not strictly exponential.
If the localization exponent $N^\mathrm{loc}$ is not energy-dependent it is clear that $\tilde{N}^\mathrm{loc}=N^\mathrm{loc}$, as it is for metallic CNTs~\cite{ComputMatSci.138.49}.
But as previously shown, the very large variations in $N^\mathrm{loc}(E)$ of the semiconducting CNTs lead to a complicated summation of conductance contributions with different exponential dependence.
In the limit of large $N$, the largest $N^\mathrm{loc}$ would dominate, but this is not a useful description, as it would be much above typical numbers of defects of experimental relevance.
With several simplifications for the Landauer-B\"uttiker formula and $N^\mathrm{loc}(E)$, solving the integral (\ref{JPC1:eqn:LBF}) yields an additional prefactor $1/(1+N)$.
But in all cases it varies not very much compared to the exponential dependence and can be treated as a slight correction of the localization exponent.
Consequently, (\ref{JPC1:eqn:Tloc}) can be used as a good approximation for estimating and predicting the conductance in the strong localization regime.

\begin{articlefigure*}[t]
	\includegraphics{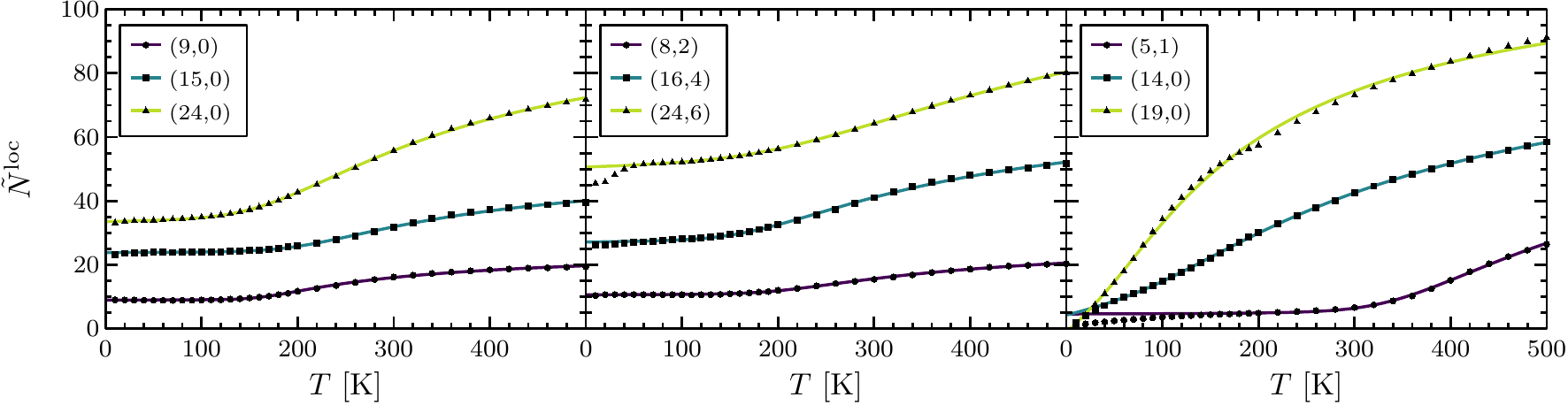}
	\caption[Localization exponent as a function of temperature for MV defects]{(color online) Temperature-dependent effective localization exponent for some examples with MV defects. The symbols are the data. The lines are regressions according to (\ref{JPC1:eqn:loc:fit}). See supplementary material for other CNTs and the regression parameters.}\label{JPC1:fig:loc(T):fit}
\end{articlefigure*}

The effective localization exponent $\tilde{N}^\mathrm{loc}$ for the conductance is calculated for $T=300\,\mathrm{K}$ and for all the different CNTs of figure \ref{JPC1:fig:CNT}.
The regressions are done around $G/G(N=1)=10^{-4}$, where the range is extended as far as possible, while keeping the regression inaccuracy within certain limits.
The results are shown in the left column of figure \ref{JPC1:fig:loc,dif(d)} for the MV (first three rows) and the DV (last row), where $\tilde{N}^\mathrm{loc}$ is depicted as a function of the tube diameter.
It shows a very good linear dependence for the four cases.
Linear regressions are depicted in figure \ref{JPC1:fig:loc,dif(d)} with straight black lines. The results are listed in table~\ref{JPC1:tab:Nloc,mfp(d)}.
As the data includes CNTs with many different chiral angles, a chirality dependence of the localization exponent can be excluded.
These result can be used as an approximate guideline to estimate and predict the conductance for CNTs with arbitrary diameters and chiral angles.
There are  deviations for smaller tube diameters, which can be explained by strong curvature effects and the resulting changes of the $\pi$-bonds.
The DV data for the $(m-n)\,\mathrm{mod}\,3=0$ case (lower row) show that $\tilde{N}^\mathrm{loc}$ is limited by a lower bound, here approximately at $d=9\,\mathrm{nm}$ and $\tilde{N}^\mathrm{loc}=5$.

As done before for the localization regime, the diffusive regime can be described by an effective elastic mean free path $\tilde{N}^\mathrm{mfp}$ at a given temperature, e.g. at $300\,\mathrm{K}$.
For the four studied cases, $\tilde{N}^\mathrm{mfp}$ is depicted as a function of the CNT diameter in the right column of figure \ref{JPC1:fig:loc,dif(d)}.
A linear trend can be seen here, too.
Linear regressions are depicted in figure \ref{JPC1:fig:loc,dif(d)} with straight black lines. The results are listed in table \ref{JPC1:tab:Nloc,mfp(d)}, too.
The deviations for small diameter tubes are likely due to curvature effects.
For the MV defect with $(m-n)\,\mathrm{mod}\,3=0$ the larger deviations are caused by the much stronger defect features near the band gap, which are not present in the other three cases.

\begin{articletable}[t]
	\begin{tabular}{l|c|c|S@{}|c|S@{}}
		{\multirow{2}{*}{defect}} & {$(m-n)$} & {$\alpha^\mathrm{loc}$} & {\multirow{2}{*}{$\beta^\mathrm{loc}$}} & {$\alpha^\mathrm{mfp}$} & {\multirow{2}{*}{$\beta^\mathrm{mfp}$}}\\[-0.3em]
		& {mod 3} & {[1/\AA]} & & {[1/\AA]} & \\
		\hline
		MV & 0 & 3.2 & -5.0 & 0.36 & -1.6  \\
		MV & 1 & 5.8 & -11  & 0.12 & -0.38 \\
		MV & 2 & 3.6 & 4.8  & 0.14 & -0.67 \\
		DV & 0 & 3.0 & -25  & 0.21 & -1.5
	\end{tabular}
	\caption[Linear regressions of $\tilde{N}^\mathrm{loc}(d)$ and $\tilde{N}^\mathrm{mfp}(d)$]{Dependence of $\tilde{N}^\mathrm{loc}$ and $\tilde{N}^\mathrm{mfp}$ on the diameter $d$ at $300\,\mathrm{K}$, described by linear regressions $\tilde{N}^\mathrm{loc}=\alpha^\mathrm{loc} d+\beta^\mathrm{loc}$ and  $\tilde{N}^\mathrm{mfp}=\alpha^\mathrm{mfp} d+\beta^\mathrm{mfp}$.}
	\label{JPC1:tab:Nloc,mfp(d)}
\end{articletable}

Figure \ref{JPC1:fig:loc(T):fit} shows some examples for the temperature dependence of $\tilde{N}^\mathrm{loc}$ for CNTs with MV defects.
The complete data can be found in the supplementary material.
The general shape of the dependence $\tilde{N}^\mathrm{loc}(T)$ can be derived under some rough assumptions and simplifications by the analytical integration of (\ref{JPC1:eqn:LBF}).
For this, the diverging shape of $N^\mathrm{loc}(E)$ is assumed to be\linebreak $N^\mathrm{loc}_0/(1-E/E_2)$.
The integral is restricted to the interval $[E_1,E_2]$.
The derivative of the Fermi distribution is approximated for large energies by $\mathrm{e}^{-(E-E_\mathrm{F})/k_\mathrm{B}T}\times\linebreak\left(1-2\mathrm{e}^{-(E-E_\mathrm{F})/k_\mathrm{B}T}\right)$.
The effective localization exponent is calculated by $\tilde{N}^\mathrm{loc} = -\frac{\partial\ln G}{\partial N}$.
Alltogether this yields a function with four free parameters, which can be used for a regression:
\begin{articleequation}
	\tilde{N}^\mathrm{loc}(T) = a + b \frac{(1-\mathrm{e}^{c-d/T})(d/T-c)}{d/T-c-(1-\mathrm{e}^{c-d/T})} \label{JPC1:eqn:loc:fit}
\end{articleequation}%
The parameters are related to the bandgap, the band edges, and the magnitude of the derivative of $N^\mathrm{loc}(E)$.
For the examples in figure \ref{JPC1:fig:loc(T):fit}, the corresponding regressions are shown as solid lines.
For different shapes, the regressions agree with the data for sufficiently high temperatures.
There are some deviations for low temperatures, where the defect-induced features in $N^\mathrm{loc}(E)$ are more dominant compared to the general trend.

\articlesection{Summary and conclusions}

We studied the influence of realistic vacancy defects on the electronic transport properties of semiconducting carbon nanotubes on a quantum level.
The influence of the vacancies is addressed by a statistical description with randomly distributed positions and orientations in CNTs with lengths up to the $\mu$m-range.
The electronic structure is described by a density-functional-based tight-binding model and a Slater-type sp$^3$-basis suitable for carbon structures.
The transport calculations are performed using quantum transport theory and linearly scaling recursive Green's function techniques to treat very large systems.
We systematically investigated a large amount of different CNTs to describe the structural dependence.
This extends previous work~\cite{NJPhys.16.123026, JComputElectron.17.521} to semiconducting CNTs with arbitrary chirality, which has not been subject of theoretical studies until now.

The strong localization regime as well as the diffusive regime are analyzed.
They are described by the dependence of the transmission as a function of the number of defects and the resulting material parameters, namely the localization length and the elastic mean free path and their dimensionless equivalents.
It is shown that a distinction concerning $(m-n)\,\mathrm{mod}\,3$ has to be made, which discriminates three groups of CNTs with different quantitative band gap dependencies.
Besides this, the localization length as well as the elastic mean free path at a given energy can be described very well by linear functions of the tube diameter, in agreement with former studies~\cite{NJPhys.16.123026, JComputElectron.17.521}.
The investigation of CNTs of different chiral angles, covering the full range, shows that a further dependence on this structural parameter can be excluded.
Furthermore, both transport regimes and the transition regime in between can be described with moderate errors by the analytical formula (\ref{JPC1:eqn:Tlocdif}) from Ref.~\cite{PhysRevB.54.5801}.

Furthermore, it was shown that the conductance within the localization and diffusive regimes can be described by effective parameters in the same way as the transmission.
Both were explicitly calculated with linear regressions for two different vacancy examples.
This description based on the conductance may be used to predict the electron transport of defective CNTs with arbitrary chirality and can help to describe CNT-based devices in micro electronics without the need of complex quantum transport computations.

\articlesection*{Acknowledgement}

This work is funded by the European Union (ERDF) and the Free State of Saxony via the ESF project\linebreak 100231947 (Young Investigators Group Computer Simulations for Materials Design - CoSiMa).
\begin{center}
	\includegraphics[width=0.4\textwidth]{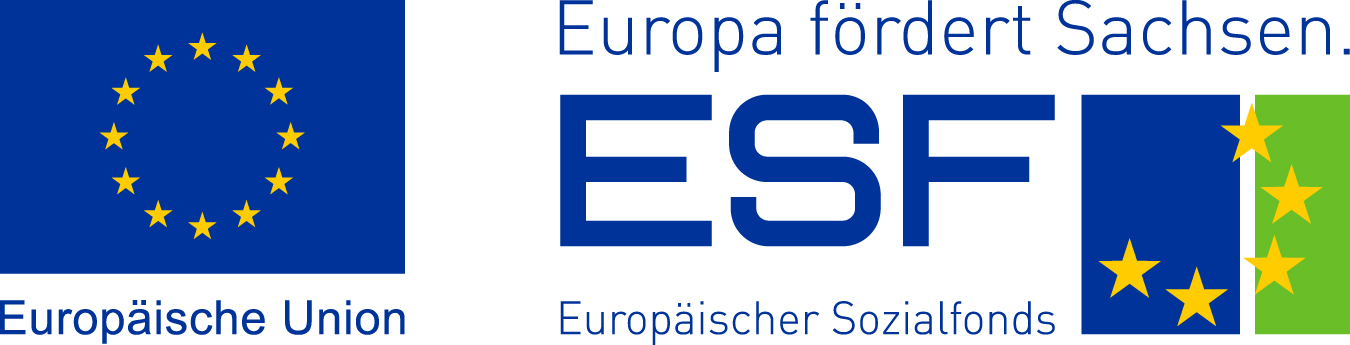}
\end{center}
The publication costs of this article were funded by the German Research Foundation/DFG and the Technische Universit\"at Chemnitz in the funding programme Open Access Publishing.

\oldonecolumn

\articlesection*{Supplementary material}
\oldaddcontentsline{toc}{section}{Supplementary material}

\begin{articlefigure*}[!h]
	\includegraphics{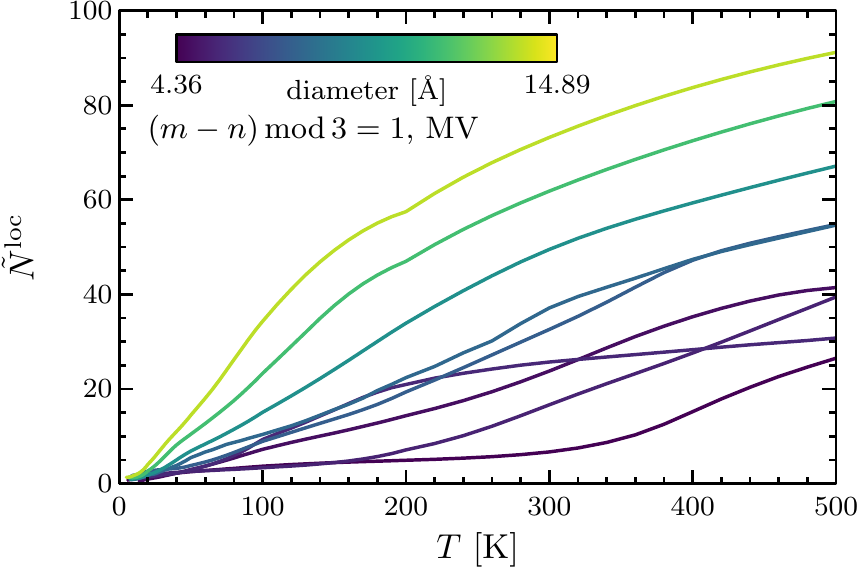}\hfill
	\includegraphics{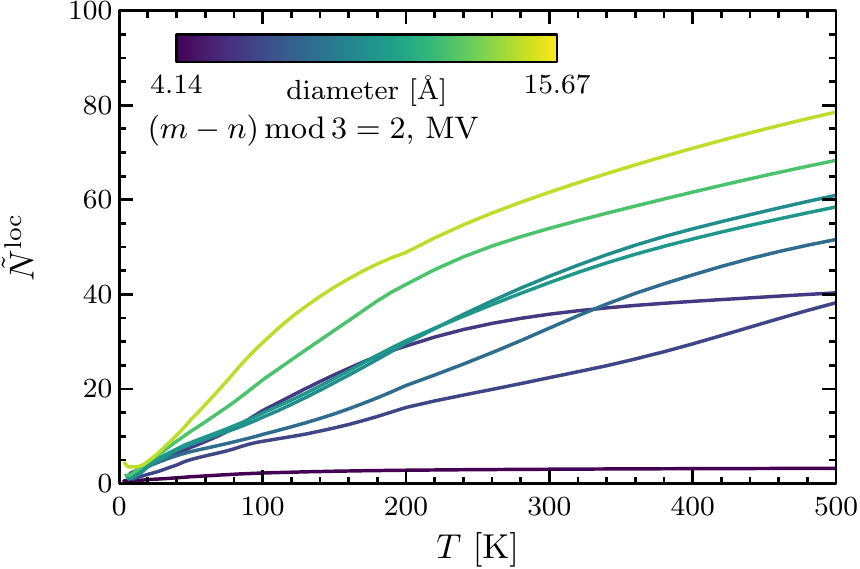}\\
	\includegraphics{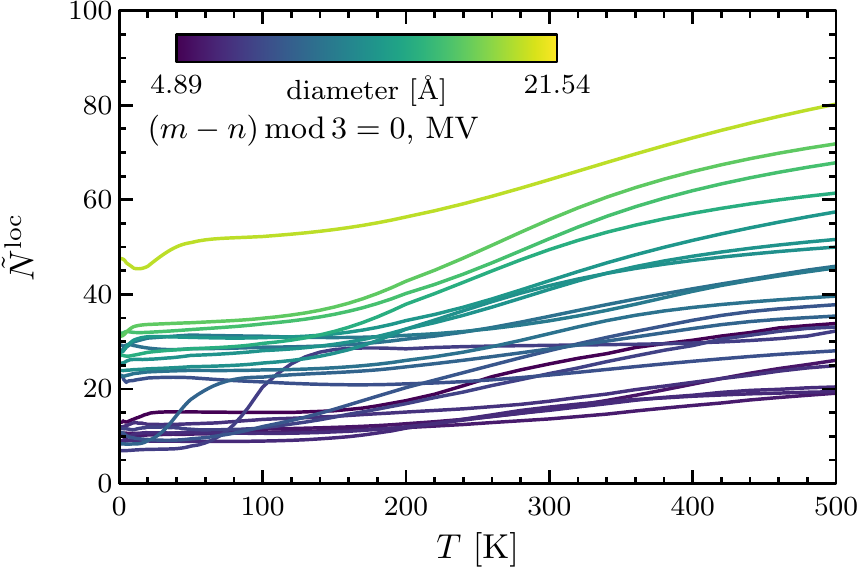}\hfill
	\includegraphics{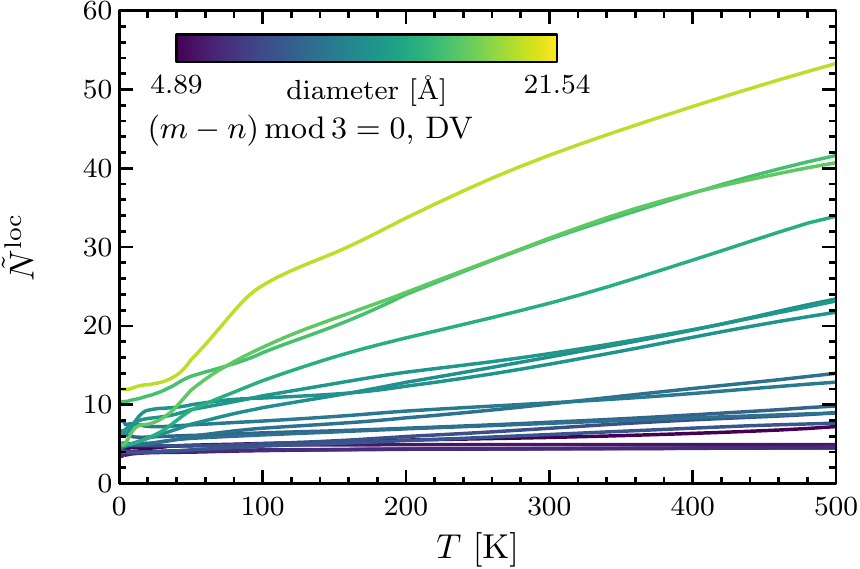}
	\caption[Localization exponents as a function of temperature]{(color online) Temperature-dependent effective localization exponent for different subsets $(m-n)\,\mathrm{mod}\,3$ and different defects (MV or DV).}
\end{articlefigure*}

\begin{articletable*}[!h]
	\centering
	\begin{tabular}{c||D{.}{.}{4}|D{.}{.}{4}|D{.}{.}{5}|D{.}{.}{5}||D{.}{.}{4}|D{.}{.}{5}|D{.}{.}{4}|D{.}{.}{5}}
		\multirow{2}{*}{CNT} & \multicolumn{4}{c||}{MV} & \multicolumn{4}{c}{DV} \\
		& \multicolumn{1}{c|}{$a$} & \multicolumn{1}{c|}{$b$} & \multicolumn{1}{c|}{$c$} & \multicolumn{1}{c||}{$d$ [eV/k$_\text{B}$]} & \multicolumn{1}{c|}{$a$} & \multicolumn{1}{c|}{$b$} & \multicolumn{1}{c|}{$c$} & \multicolumn{1}{c}{$d$ [eV/k$_\text{B}$]} \\
		\hline
		(9,0)  &    8.114  &   0.8829 & 19.43    & 0.2695  &    3.213  &   0.6668  &  0.0    & 0.01163 \\
		(12,0) &   19.37   &   1.345  & 13.74    & 0.3146  &    2.998  &   1.792   &  5.273  & 0.1159  \\
		(15,0) &   21.52   &   2.353  & 13.52    & 0.2425  &  -21.26   &  26.2     &  0.0    & 0.1501  \\
		(18,0) &   20.89   &   3.696  & 11.65    & 0.1587  & -767.3    & 773.6     &  3.278  & 2.146   \\
		(21,0) &   21.45   &   6.725  &  9.123   & 0.1379  &  -61.22   &  65.95    &  0.0    & 0.1122  \\
		(24,0) &   24.38   &   9.176  &  8.449   & 0.1456  &  -49.27   &  55.38    &  0.0    & 0.06249 \\
		(7,0)  &  -17.22   &  19.19   &  6.387   & 0.1878  &           &           &         &         \\
		(10,0) &  -72.22   &  73.37   &  1.748   & 0.1123  &           &           &         &         \\
		(13,0) &  -88.62   &  88.26   &  0.6711  & 0.06348 &           &           &         &         \\
		(16,0) & -107.0    & 106.2    &  0.2528  & 0.04695 &           &           &         &         \\
		(19,0) & -116.1    & 114.9    &  0.1168  & 0.03578 &           &           &         &         \\
		(8,0)  &   -8.22   &  12.45   &  4.48    & 0.0409  &           &           &         &         \\
		(11,0) &  -40.86   &  45.57   &  2.491   & 0.1016  &           &           &         &         \\
		(14,0) &  -56.18   &  60.46   &  1.251   & 0.06808 &           &           &         &         \\
		(17,0) &  -94.23   &  95.04   &  0.0     & 0.04585 &           &           &         &         \\
		(20,0) &  -98.41   &  99.24   &  0.04263 & 0.03772 &           &           &         &         \\
		(10,1) &    6.335  &   0.7693 & 38.21    & 0.184   &           &           &         &         \\
		(7,1)  &    2.005  &   8.937  &  6.991   & 0.2292  &    4.702  &   0.01742 & 15.1    & 0.04319 \\
		(14,2) &   22.54   &   5.775  &  8.258   & 0.1975  &   -0.6879 &   5.753   &  0.0    & 0.05928 \\
		(5,1)  &    0.6108 &   4.026  & 19.84    & 0.5764  &           &           &         &         \\
		(8,2)  &    9.215  &   1.441  & 13.39    & 0.2372  &   -0.6039 &   4.679   &  0.0    & 0.4103  \\
		(12,3) &   13.83   &   7.972  &  5.085   & 0.1439  &   -2.125  &   8.281   &  5.958  & 0.3732  \\
		(16,4) &   21.48   &   5.688  &  8.848   & 0.1547  &  -53.94   &  60.52    &  0.4481 & 0.2022  \\
		(20,5) &   19.77   &  12.16   &  6.788   & 0.1367  &  -54.45   &  64.63    &  0.0    & 0.09628 \\
		(24,6) &   32.81   &  17.85   &  5.116   & 0.1528  &  -15.57   &  27.3     &  2.382  & 0.05215 \\
		(6,2)  &   -5.697  &   6.851  &  6.261   & 0.04993 &           &           &         &         \\
		(9,3)  &    5.102  &   6.404  &  7.297   & 0.1376  &           &           &         &         \\
		(5,2)  &   13.06   &   1.967  & 17.07    & 0.2816  &  -24.92   &  29.69    &  5.073  & 0.7881  \\
		(10,4) &   -9.403  &  17.74   &  3.499   & 0.08113 &   -4.822  &   8.764   &  0.0    & 0.1161  \\
		(15,6) &   25.57   &   5.202  & 10.98    & 0.2136  &  -35.56   &  42.25    &  0.0    & 0.133   \\
		(4,2)  &   -1.083  &   1.611  &  1.96    & 0.01568 &           &           &         &         \\
		(6,3)  &    1.942  &   8.355  &  4.648   & 0.1914  &    2.647  &   0.9867  &  0.0    & 0.01515 \\
		(8,4)  &  -99.68   &  99.81   &  0.484   & 0.09817 &           &           &         &         \\
		(10,5) & -109.3    & 110.2    &  0.0     & 0.07889 &           &           &         &         \\
		(12,6) &   28.67   &   2.364  & 16.37    & 0.3963  &  -40.77   &  46.73    &  0.1084 & 0.3168  \\
		(7,4)  &  -18.23   &  30.31   &  2.956   & 0.2394  &           &           &         &         \\
		(6,4)  &  -94.78   &  96.12   &  0.0     & 0.1364  &           &           &         &         \\
		(4,3)  & -108.2    & 107.9    &  0.6031  & 0.1505  &           &           &         &
	\end{tabular}
	\caption[Regression parameters of $\tilde{N}^\mathrm{loc}(d)$ and $\tilde{N}^\mathrm{mfp}(d)$]{Parameters $a$, $b$, $c$, and $d$ for the regression $\tilde{N}^\mathrm{loc}(T) = a + b(1-\mathrm{e}^{c-d/T})(d/T-c)/[d/T-c-(1-\mathrm{e}^{c-d/T})]$.}
\end{articletable*}

\end{document}